\documentclass[%
 reprint,
 amsmath,amssymb,
 aps,
 prc,
 groupedaddress,
 floatfix,
]{revtex4-2}

\usepackage{graphicx}
\usepackage{dcolumn}
\usepackage{bm}
\usepackage{comment}
\usepackage{color}
\usepackage{setspace} 
\usepackage{multirow}
\usepackage{tablefootnote}
\usepackage{threeparttable}

\newcommand{\tohoku}{$^1$}
\newcommand{\obihiro}{$^2$}
\newcommand{\ipmu}{$^3$}
\newcommand{\gppu}{$^4$}
\newcommand{\fris}{$^5$}
\newcommand{\tohokuRigaku}{$^6$}
\newcommand{\osaka}{$^7$} 
\newcommand{\osakarcnp}{$^8$}
\newcommand{\tokushima}{$^9$}
\newcommand{\tokushimaGakusei}{$^10$}
\newcommand{\lbl}{$^{11}$}
\newcommand{\hawaii}{$^{12}$}
\newcommand{\mituniv}{$^{13}$}
\newcommand{\tennessee}{$^{14}$}
\newcommand{\tunl}{$^{15}$}
\newcommand{\chapehill}{$^{16}$}
\newcommand{\northcarolina}{$^{17}$}
\newcommand{\duke}{$^{18}$}
\newcommand{\virginia}{$^{19}$}
\newcommand{\seattle}{$^{20}$}
\newcommand{\nikhef}{$^{21}$}
\newcommand{\bu}{$^{22}$}

\newcommand{\tohokuaff}{\affiliation{\tohoku Research Center for Neutrino Science, Tohoku University, Sendai 980-8578, Japan}}
\newcommand{\ipmuaff}{\affiliation{\ipmu Institute for the Physics and Mathematics  of the Universe, The University of Tokyo, Kashiwa 277-8568, Japan}}
\newcommand{\gppuaff}{\affiliation{\gppu Graduate Program on Physics for the Universe, Tohoku University, Sendai 980-8578, Japan}}
\newcommand{\frisaff}{\affiliation{\fris Frontier Research Institute for Interdisciplinary Sciences, Tohoku University, Sendai 980-8578, Japan}}
\newcommand{\tohokuRigakuaff}{\affiliation{\tohokuRigaku Department of Physics, Tohoku University, Sendai 980-8578, Japan}}
\newcommand{\osakarcnpaff}{\affiliation{\osakarcnp Research Center for Nuclear Physics (RCNP), Osaka University, Ibaraki, Osaka 567-0047, Japan}}
\newcommand{\osakaaff}{\affiliation{\osaka Graduate School of Science, Osaka University, Toyonaka, Osaka 560-0043, Japan}}
\newcommand{\tokushimaaff}{\affiliation{\tokushima Graduate School of Advanced Technology and Science, Tokushima University, Tokushima 770-8506, Japan}}
\newcommand{\tokushimaGakuseiaff}{\affiliation{\tokushimaGakusei Graduate School of Integrated Arts and Sciences, Tokushima University, Tokushima 770-8502, Japan}}
\newcommand{\lblaff}{\affiliation{\lbl Nuclear Science Division, Lawrence Berkeley National Laboratory, Berkeley, CA 94720, USA}}
\newcommand{\hawaiiaff}{\affiliation{\hawaii Department of Physics and Astronomy, University of Hawaii at Manoa, Honolulu, HI 96822, USA}}
\newcommand{\mitunivaff}{\affiliation{\mituniv Massachusetts Institute of Technology, Cambridge, MA 02139, USA}}
\newcommand{\buaff}{\affiliation{\bu Boston University, Boston, MA 02215, USA}}
\newcommand{\tennesseeaff}{\affiliation{\tennessee Department of Physics and Astronomy,  University of Tennessee, Knoxville, TN 37996, USA}}
\newcommand{\tunlaff}{\affiliation{\tunl Triangle Universities Nuclear Laboratory, Durham, NC 27708, USA}}    
\newcommand{\chapehillaff}{\affiliation{\chapehill The University of North Carolina at Chapel Hill, Chapel Hill, NC 27599, USA}}
\newcommand{\northcarolinaaff}{\affiliation{\northcarolina North Carolina Central University, Durham, NC 27701, USA}}
\newcommand{\dukeaff}{\affiliation{\duke Physics Department at Duke University, Durham, NC 27705, USA}}
\newcommand{\seattleaff}{\affiliation{\seattle Center for Experimental Nuclear Physics and Astrophysics, University of Washington, Seattle, WA 98195, USA}}
\newcommand{\nikhefaff}{\affiliation{\nikhef Nikhef and the University of Amsterdam, Science Park,  Amsterdam, the Netherlands}}
\newcommand{\virginiaaff}{\affiliation{\virginia Center for Neutrino Physics, Virginia Polytechnic Institute and State University, Blacksburg, VA 24061, USA}}
\newcommand{\obihiroaff}{\affiliation{\obihiro Department of Human Science, Obihiro University of Agriculture and Veterinary Medicine, Obihiro, Hokkaido 080-8555, Japan}}

\begin{document}


\title{Measurement of cosmic-ray muon spallation products in a xenon-loaded liquid scintillator with KamLAND}

\author{S.~Abe\tohoku}
\author{S.~Asami\tohoku}
\author{M.~Eizuka\tohoku}
\author{S.~Futagi\tohoku}
\author{A.~Gando\tohoku}
\author{Y.~Gando\tohoku$^,$\obihiro} 
\author{T.~Gima\tohoku}
\author{A.~Goto\tohoku}
\author{T.~Hachiya\tohoku}
\author{K.~Hata\tohoku}
\author{K.~Hosokawa\tohoku}\altaffiliation[Present address: ]{Kamioka Observatory, Institute for Cosmic Ray Research, The University of Tokyo, Higashi-Mozumi, Kamioka, Hida, Gifu 506-1205, Japan}
\author{K.~Ichimura\tohoku}
\author{S.~Ieki\tohoku}
\author{H.~Ikeda\tohoku}
\author{K.~Inoue\tohoku$^,$\ipmu} 
\author{K.~Ishidoshiro\tohoku}
\author{Y.~Kamei\tohoku}\altaffiliation[Present address: ]{RIKEN, Wako, Saitama 351-0198, Japan}
\author{N.~Kawada\tohoku}
\author{Y.~Kishimoto\tohoku$^,$\ipmu}
\author{M.~Koga\tohoku$^,$\ipmu} 
\author{M.~Kurasawa\tohoku}
\author{T.~Mitsui\tohoku}
\author{H.~Miyake\tohoku}
\author{T.~Nakahata\tohoku}
\author{K.~Nakamura\tohoku}\altaffiliation[Present address: ]{Faculty of Health Sciences, Butsuryo College of Osaka, Sakai, Osaka 593-8328, Japan}  
\author{R.~Nakamura\tohoku}
\author{H.~Ozaki\tohoku$^,$\gppu}
\author{T.~Sakai\tohoku}
\author{I.~Shimizu\tohoku}
\author{J.~Shirai\tohoku}
\author{K.~Shiraishi\tohoku}
\author{A.~Suzuki\tohoku}
\author{Y.~Suzuki\tohoku} 
\author{A.~Takeuchi\tohoku} \altaffiliation[Present address: ]{Faculty of Science, The University of Tokyo, Bunkyo-ku, Tokyo 113-0033, Japan}
\author{K.~Tamae\tohoku}
\author{H.~Watanabe\tohoku}
\author{Y.~Yoshida\tohoku}
\author{S.~Obara\fris} \altaffiliation[Present address: ]{National Institutes for Quantum Science and Technology, Sendai, 980-8579, Japan}
\author{A.~K.~Ichikawa\tohokuRigaku}
\author{S.~Yoshida\osaka}
\author{S.~Umehara\osakarcnp}
\author{K.~Fushimi\tokushima}
\author{K.~Kotera\tokushima}
\author{Y.~Urano\tokushima}
\author{B.~E.~Berger\lbl$^,$\ipmu}
\author{B.~K.~Fujikawa\lbl$^,$\ipmu}
\author{J.~G.~Learned\hawaii}
\author{J.~Maricic\hawaii}
\author{S.~N.~Axani\mituniv}
\author{Z.~Fu\mituniv}
\author{J.~Smolsky\mituniv}
\author{L.~A.~Winslow\mituniv}
\author{Y.~Efremenko\tennessee$^,$\ipmu}
\author{H.~J.~Karwowski\tunl$^,$\chapehill}
\author{D.~M.~Markoff\tunl$^,$\northcarolina}
\author{W.~Tornow\tunl$^,$\duke$^,$\ipmu}
\author{S.~Dell'Oro\virginia}
\author{T.~O'Donnell\virginia}
\author{J.~A.~Detwiler\seattle$^,$\ipmu}
\author{S.~Enomoto\seattle$^,$\ipmu}
\author{M.~P.~Decowski\nikhef$^,$\ipmu}
\author{K.~Weerman\nikhef}
\author{C.~Grant\bu}
\author{A.~Li\bu$^,$\chapehill}
\author{H.~Song\bu}

\collaboration{KamLAND-Zen Collaboration}

\tohokuaff 
\obihiroaff
\ipmuaff
\gppuaff
\frisaff
\tohokuRigakuaff
\osakaaff
\osakarcnpaff
\tokushimaaff
\tokushimaGakuseiaff
\lblaff
\hawaiiaff
\mitunivaff
\tennesseeaff
\tunlaff
\chapehillaff
\northcarolinaaff
\dukeaff
\virginiaaff
\seattleaff
\nikhefaff
\buaff
\date{\today}

\begin{abstract}
Cosmic-ray muons produce various radioisotopes when passing through material. 
These spallation products can be backgrounds for rare event searches such as in solar neutrino, double-beta decay, and dark matter search experiments.
The KamLAND-Zen experiment searches for neutrinoless double-beta decay in 745\,kg of xenon dissolved in liquid scintillator.
The experiment includes dead-time-free electronics with a high efficiency for detecting muon-induced neutrons.
The production yields of different radioisotopes are measured with a combination of delayed coincidence techniques, newly developed muon reconstruction and xenon spallation identification methods.
The observed xenon spallation products are consistent with results from the FLUKA and Geant4 simulation codes.
\end{abstract}

\maketitle

\section{\label{sec:intro}Introduction}
Cosmic-ray muons generate radioisotopes with decay products that can be critical backgrounds for rare event experiments.
Experiments searching for solar neutrinos~\cite{superK_solar,SNO_solar}, neutrinoless double-beta ($0\nu\beta\beta$) decay~\cite{Zen800,EXO-200} or dark matter interactions~\cite{xenon1t} may suffer from these backgrounds.
Muons can be suppressed by locating the detector underground, for instance, the KamLAND-Zen~\cite{Zen400,Zen800} experiment is sited at a depth of 2700\,m-water-equivalent, reducing the muon rate passing the detector to 0.34\,Hz~\cite{spallation2010}.
Nevertheless, the remaining muon flux can induce spallation interactions in the detector material. 

The influence of spallation backgrounds can be reduced by vetoing the detector for a short time after a muon passes through, depending on  
the lifetimes of the produced isotopes. 
Since the impact on background estimates also depend on the $Q$-values, the understanding of isotope production is crucial. 
Liquid scintillator (LS) or water Cherenkov detectors mainly consist of relatively light isotopes such as $^{12}$C and $^{16}$O, but $0\nu\beta\beta$ decay and direct detection dark matter experiments may include much heavier isotopes. 
Heavy isotopes can induce a larger variety of spallation products with long radioactive decay chains.
Knowledge of the characteristics of the events generated by spallation processes such as the correlation of the radioisotope with the muon track and the detailed particle emission are important for background discrimination techniques.
Various spallation studies have been reported from underground experiments~\cite{superK_shower, borexino, KamNet}. 
One of the core challenges in this work is the efficient identification of spallation products.
Muons crossing the LS leave a very large signal affecting the readout electronics and require a large dynamic range in the data acquisition system. 
Some spallation backgrounds, such as the small signals from capture of muon-induced neutrons after the muon passing, may be affected by these readout effects.

Cross section uncertainties impact the spallation background estimate in the energy spectrum.
We compare our measurements to the energy spectra for spallation backgrounds which are reconstructed using Monte Carlo simulations (FLUKA~\cite{FLUKA1, FLUKA2} and Geant4~\cite{Geant4_1, Geant4_2, Geant4_3}).

We describe the measurement of muon-induced isotopes produced in KamLAND.
The data was acquired from Feb. 5th, 2019 to May 8th, 2021 and includes the $0\nu\beta\beta$ search period using 745\,kg of xenon~\cite{Zen800}.
This paper is structured as follows.
The experimental setup and the data acquisition system are introduced in Sec.~\ref{sec:detector} and the event reconstruction is reported in Sec.~\ref{sec:EvRecon}. 
Results from FLUKA and Geant4, the Monte Carlo (MC) programs used to simulate spallation backgrounds, are discussed in Sec.~\ref{sec:mc}, followed by the spallation background measurements in Sec.~\ref{sec:data} and we summarize in Sec.~\ref{sec:summary}.

\section{\label{sec:detector}Detector}
\begin{figure}
    \includegraphics[width=1.0\linewidth]{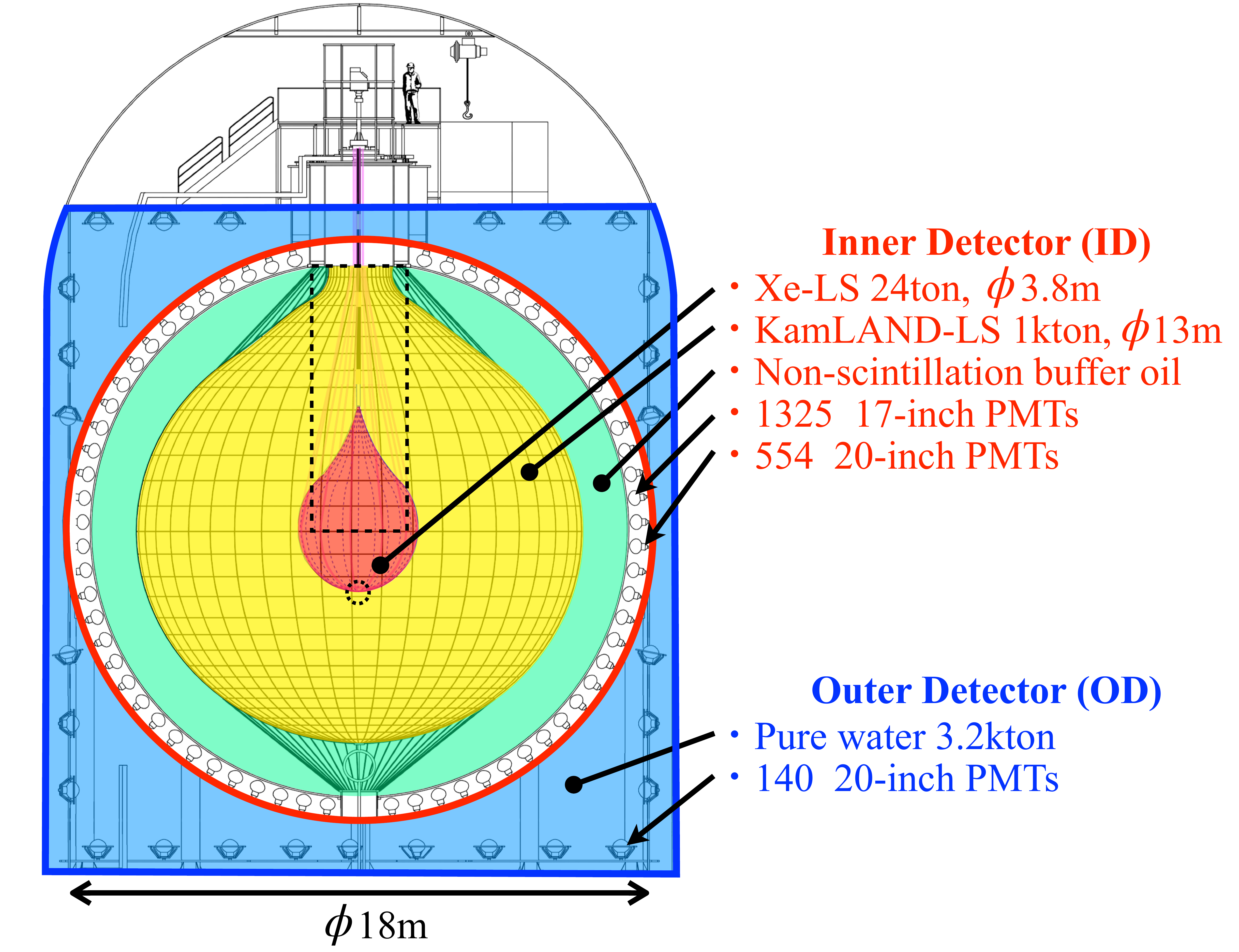}
    \caption{\label{fig:kamland} Schematic view of the KamLAND-Zen detector. The black dotted rectangle, corresponding to a 1.5\,m-radius cylinder in the upper hemisphere, illustrates a tube cut that is applied to exclude the droplet-like region of Xe-LS. 
    The 0.7\,m-diameter sphere centered at $(x,y,z)=(0,0,-1.9\,$m) is vetoed due to contamination (hot spot veto).}
\end{figure}

The Kamioka Liquid scintillator Anti-Neutrino Detector (KamLAND) is located about 1000\,m below the peak of Mt. Ikenoyama, Gifu, Japan.
The experiment contains an inner scintillation detector and an outer water Cherenkov detector (see Fig.~\ref{fig:kamland}).
The inner detector (ID) consists of a 18\,m-diameter stainless-steel spherical tank with a 13\,m-diameter nylon-EVOH balloon at its center.
The balloon is filled with 1\,kton of LS (KamLAND-LS).
The KamLAND-LS is a composition of dodecane (0.753\,g/cm$^{3}$), pseudocumene (0.875\,g/cm$^{3}$) which is also called 1,2,4-trimethylbenzene or PC, and PPO (2,5-diphyenyloxazole) as the fluor (see Table~\ref{tab:LScomponent}).
The density of the KamLAND-LS is 0.780\,g/cm$^{3}$ at 11.5$^\circ$C. 
The remaining volume outside of the balloon is filled with non-scintillating buffer oil.
The buffer oil is a mixture of 57\% isoparaffin and 43\% dodecane by volume.
The scintillation light output is about 8000\,photons/MeV, which is observed by 1325 17-inch photomultiplier tubes (PMTs) and 554 20-inch PMTs bolted to the inner surface of the stainless-steel sphere. 
The total photocathode coverage is 34\%.

For the $0\nu\beta\beta$ decay search,  a 3.8\,m-diameter inner balloon (IB)~\cite{miniballoon}  was installed on May 9th 2018, containing 30.5$\pm$0.3\,m$^3$ of xenon-loaded LS (Xe-LS).
The composition of Xe-LS is decane (0.731\,g/cm$^{3
}$), PC (0.875\,g/cm$^{3}$) and PPO as shown in Table~\ref{tab:LScomponent}.
The KamLAND-LS is 10\% brighter than Xe-LS.
A total of 745\,kg of xenon is dissolved in this scintillator.
The xenon components of the Xe-LS are given in Table~\ref{tab:xenon}.
\begin{table}[h]
\caption{\label{tab:LScomponent} Components of the two liquid scintillators used in KamLAND-Zen.}
\begin{ruledtabular}
\begin{tabular}{lcc}
\textrm{Material}&
\textrm{KamLAND-LS}&
\textrm{Xe-LS} \\
\colrule
Dodecane (vol\%) & 80.2 & - \\
Decane (vol\%) & -  & 82.4\\
PC (vol\%) & 19.8 & 17.6\\
PPO (mg/cm$^{3}$) & 1.36$\,\pm\,$0.03 & 2.38$\,\pm\,$0.02\\
Xe (wt\%) & - & 3.13$\,\pm\,0.01$\\
\end{tabular}
\end{ruledtabular}
\caption{\label{tab:xenon} Xenon composition in the Xe-LS, based on the quality specification sheet.}
\begin{ruledtabular}
\begin{tabular}{lccc}
\textrm{}&
\textrm{Atomic mass\,(u)}&
\textrm{Volume ratio\,(\%)}&
\textrm{Mass\,(kg)}\\
\colrule
$^{136}$Xe & 135.907~\cite{xenonmass} & 90.85 & 677.4\\
$^{134}$Xe & 133.905 & 8.82 & 64.8\\
$^{132}$Xe & 131.904 & 0.17 & 1.3 \\
Others & - & 0.16 & 1.5\\
\hline
Total & 135.80 & 100.00 & 745.0\\
\end{tabular}
\end{ruledtabular}
\end{table}

The outer detector (OD) is a shield for $\gamma$-rays and fast neutrons from the surrounding rock.
It is a 20\,m-diameter and 20\,m-high cylindrical cavern filled with 3.2\,kton of pure water. 
Cosmic-ray muons are identified by detecting Cherenkov light with 20-inch PMTs, including ones with high quantum efficiency~\cite{ODref}.

PMT waveforms are digitized by two separate data acquisition (DAQ) systems.
The KamLAND Front-End Electronics (KamFEE) system has been working as the main DAQ since the beginning of KamLAND in 2002.
The other system, Module for General-Use Rapid Application (MoGURA), was installed in Aug. 2010 and plays an essential role in tagging muon-induced neutrons.
These neutrons are quickly thermalized with a mean capture time of 207.5\,$\mu$s, and can be identified by a 2.2\,MeV capture $\gamma$-ray on $^{1}$H~\cite{spallation2010}.

The KamFEE system samples PMT signals with the Analog Transient Waveform Digitizer (ATWD) from the 17-inch and 20-inch PMTs.
Three amplifier gains ($\times\,0.5, \times\,4, \times\,20$) cover a wide dynamic range from 1 photo-electron (p.e.) to 1000\,p.e. and the waveform from each gain is digitized.
An ATWD stores 128 samples with 10-bit resolution and a sampling interval of 1.5\,ns.
Since the analog-to-digital conversion takes 27\,$\mu$s, two ATWDs are assigned to each PMT in order to avoid potential dead-time.
Nevertheless, the high event rate after muon-induced events and a baseline distortion of the PMT signal introduce a significant amount of dead-time in the KamFEE electronics following muon events.

The MoGURA system was designed to be a dead-time-free DAQ system.
Data is read out from 17-inch PMTs only. 
Unlike KamFEE, MoGURA implements a 1\,GHz sampling 8-bit fast-ADC which is connected to a $\times\,$120 gain amplifier, and three 200\,MHz 8-bit sampling pipeline-ADCs each connected to separate amplifiers gains ($\times\,24, \times\,2.4, \times\,0.24$).
The dynamic range of 100,000 covers from 1\,p.e. to the cosmic-ray muon signal.

\begin{figure}
    \includegraphics[width=1.0\linewidth]{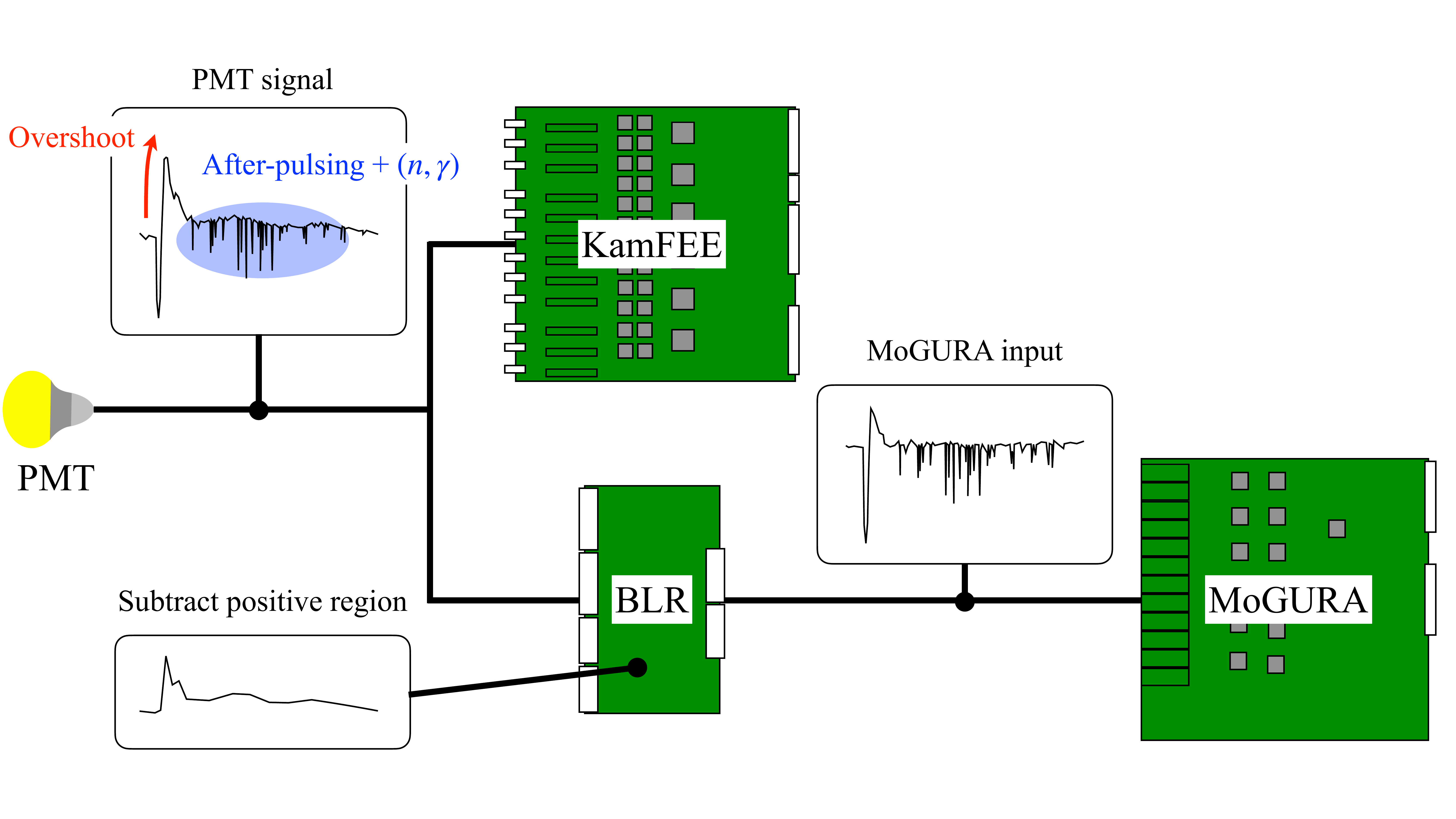}
    \caption{\label{fig:BLR}Schematic of KamFEE, BLR and MoGURA. After a high charge event, the PMT baseline overshoot is subtracted and the baseline stabilized.} 
\end{figure}

\begin{figure}
    \includegraphics[width=0.7\linewidth]{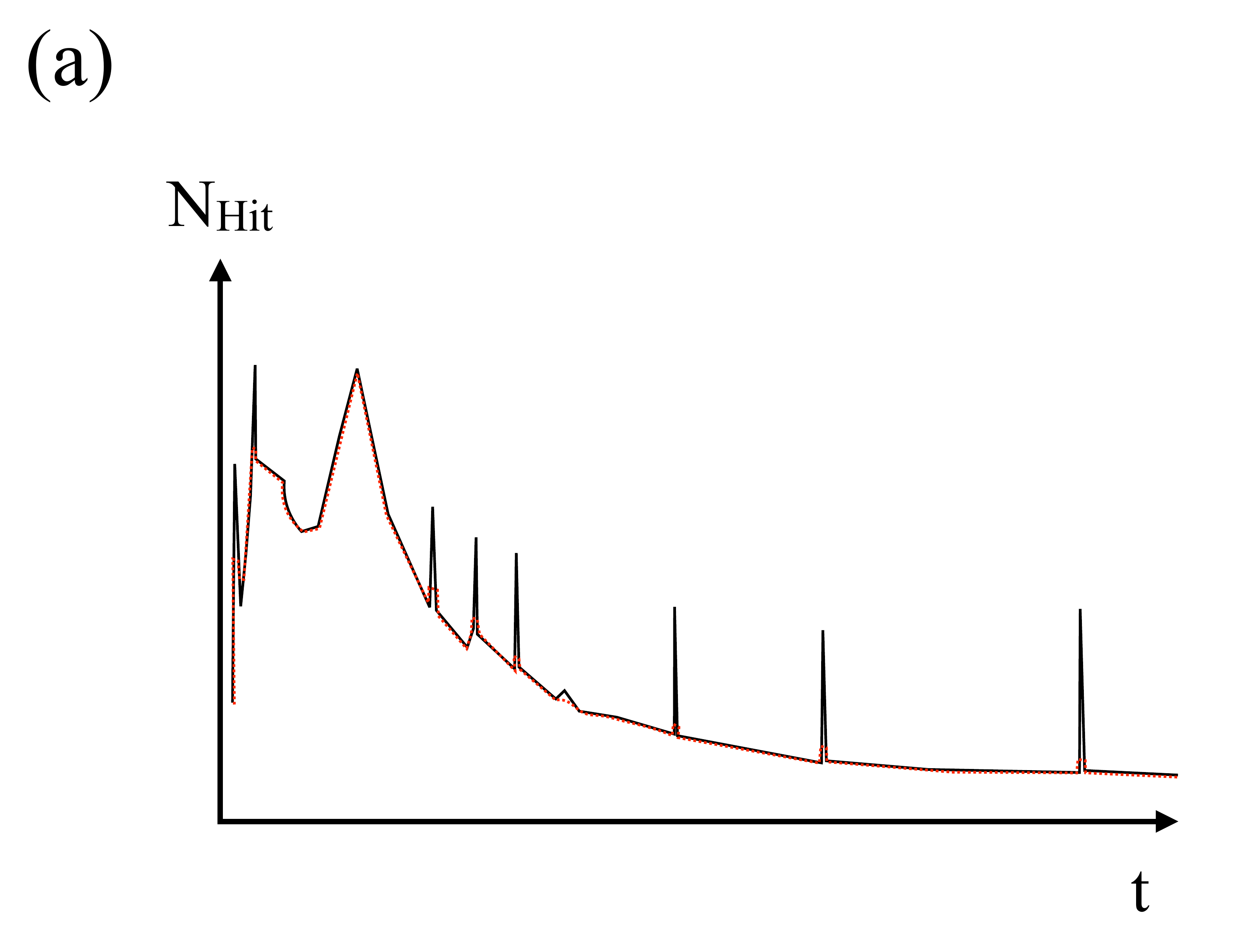}
    \includegraphics[width=0.7\linewidth]{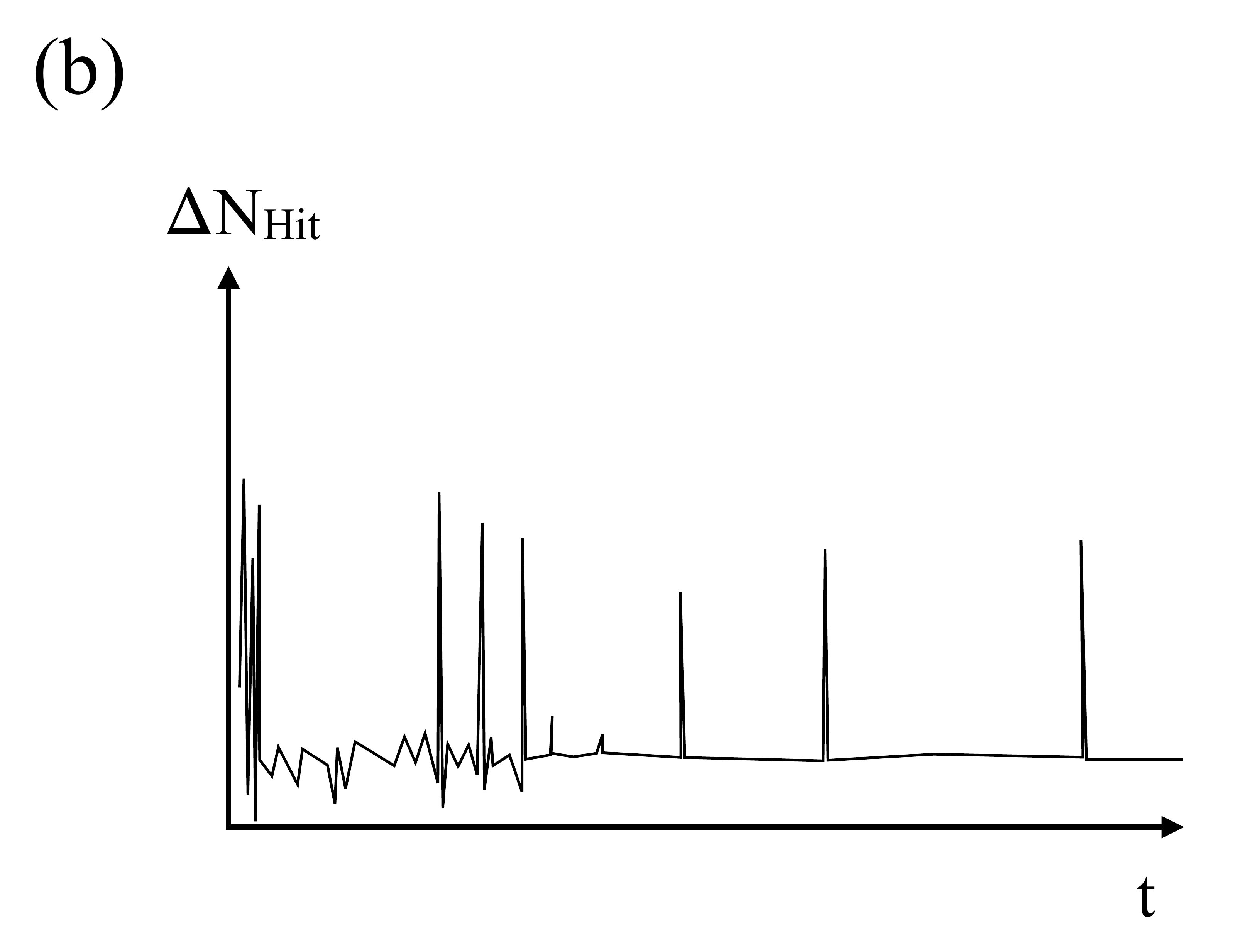}
    \caption{\label{fig:adaptive}Illustration of the adaptive trigger in MoGURA. (a) Number of PMT hits in a 120\,ns window following a muon crossing. The red dashed line shows the average of $N_{Hit}$ for 240\,ns. (b) The adaptive trigger uses the difference of $N_{Hit}$ and its 240-ns average to efficiently extract the sharp peaks from neutron capture events.
    }
\end{figure}

When high charge events occur, such as from muons, the PMT baseline is distorted for $\mathcal{O}$(1\,ms).
Small neutron capture signals follow shortly later ($<\mathcal{O}$($100\,\mu$s)), so the neutron capture signals cannot be easily identified.
The neutron tagging is enabled by a baseline restorer (BLR) and MoGURA's so-called adaptive mode.
The BLR is an electric circuit which contributes to reducing the dead-time.
The signal from each PMT is divided between a KamFEE channel and BLR (see Fig.~\ref{fig:BLR}).
The BLR stabilizes the baseline by subtracting the overshoot region and the signal is sent to the MoGURA board.
The MoGURA trigger system is continuously calculating the total number of hits ($N_{Hit}$) from 17-inch PMTs in a sliding time-window of 120\,ns.
A special trigger is launched if $N_{Hit}$ exceeds a threshold of $\sim$\,800 and MoGURA is switched to the adaptive mode for 1\,ms.
In that condition, 
$\Delta N_{Hit}$, defined as $N_{Hit}$ subtracted by its 240\,ns average, is calculated.
Based on the $\Delta N_{Hit}$ value, adaptive triggers are issued to record neutron capture events (see Fig.~\ref{fig:adaptive}).
The adaptive trigger helps discriminating signals from artifacts such as PMT after-pulsing and ringing.

\section{\label{sec:EvRecon}Event reconstruction}
Cosmic-ray muons traversing the detector may be followed by neutron captures, $\beta$ decays and $\alpha$ decays of the spallation isotopes.
The time-space correlations between the muons and subsequent events are key for our spallation identification techniques. 
Recent simulation~\cite{superK_shower} reports that the $\beta$ decay isotopes are produced by muon-induced shower secondaries, rather than the cosmic-ray muons themselves, so the energy deposit of the secondaries reflect the position of spallation reactions.
These aspects were taken into account by introducing track and shower position information to our delayed coincidence methods (see Sec.~\ref{sec:data}).
The space correlation between the $\beta$ decay and neutron capture is also included in our methods. 
This section describes the energy calibration, cosmic-ray muon related parameters (identification, track, shower), and vertex reconstruction of neutron capture.

\subsection{\label{sec:EnergyCalib}Energy calibration}
The KamLAND experiment was calibrated with the radioactive sources listed in Table~\ref{tab:calib}.
The sources were deployed at various positions within 5.5\,m from the detector center by introducing an off-axis calibration system~\cite{calib}.
In addition, neutrons and $^{12}$B from spallation reactions were utilized.
$^{12}$B ($\tau_{1/2}=20.2$\,ms, $Q=13.4$\,MeV) is produced uniformly in the detector and its $\beta^{-}$ decay spectrum provides a high-energy calibration source.
Neutron captures on protons produce a monochromatic $\gamma$-ray at 2.2\,MeV which is uniformly distributed in the LS.
We correct the non-linear relation between the visible energy and the deposited energy from $\beta$-rays and $\gamma$-rays with a phenomenological model based on Birks quenching~\cite{birks1}.
\begin{table}[h]
\caption{\label{tab:calib} Calibration sources used in the KamLAND experiment~\cite{calib}.}
\begin{ruledtabular}
\begin{tabular}{cc}
\textrm{Source}&
\textrm{$\gamma$\,(MeV)}\\
\colrule
$^{203}$Hg & 0.279\\
$^{137}$Cs & 0.662\\
$^{68}$Ge & 2$\,\times\,$0.511\\
$^{65}$Zn & 1.116\\
$^{60}$Co & 1.173+1.333\\
\colrule
$^{241}$Am-$^9$Be & 2.22\footnote{$^{1}$H + $n\,\rightarrow\,^{2}$H + $\gamma$ (2.22\,MeV)}, (4.44, 7.65, 9.64)\footnote{$^9$Be + $\alpha\,\rightarrow\,^{13}$C$^{*}\,\rightarrow\,^{12}$C + $n + \gamma s$}\\
\end{tabular}
\end{ruledtabular}
\end{table}

\subsection{\label{sec:muon}Cosmic-ray muon identification}
Cosmic-ray muons are identified by measuring the large amount of scintillation and Cherenkov light associated with muons crossing the LS ($LS \ muons$).
The selection criteria are defined by the total number of photo-electrons observed by the 17-inch PMTs in the ID ($Q_{ID}$), where $Q_{ID}>40000$\,p.e. is an indication for a muon event.

\subsection{\label{sec:mutrack}Cosmic-ray muon track}
\begin{figure}
    \includegraphics[width=1.\linewidth]{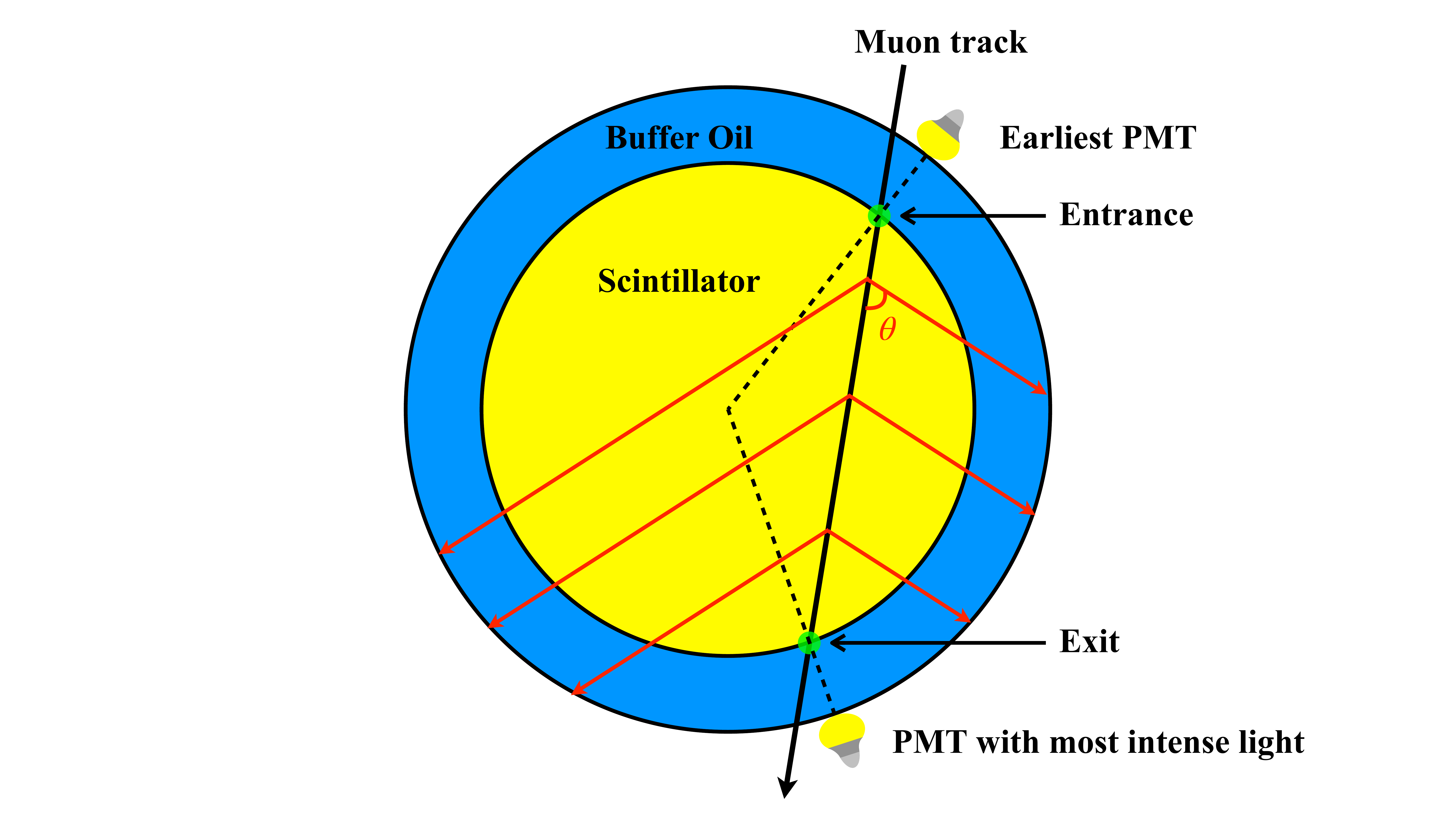}
    \caption{\label{fig:TrackReconstruction} Schematic of the main muon track reconstruction algorithm. The entrance point is determined by the earliest hit PMT, while the muon exit point is determined by the PMT with the largest signal. The earliest scintillation light and Cherenkov light propagate with Cherenkov angle $\theta$.}
\end{figure}

A muon track is reconstructed by finding the detector entrance and exit points.
The photons arriving first at the PMT determine the entrance point.
For relativistic muons, both the earliest scintillation light and the Cherenkov light are emitted along the muon track with the Cherenkov angle, so that the muon exit is expected to be near the PMT which observes the most intense light.
By connecting the earliest hit PMT (largest signal PMT) and the center of the detector, as shown in Fig.\,\ref{fig:TrackReconstruction}, the entrance (exit) is found at the intersection with the balloon surface.
The tilt and position of the reconstructed track are corrected to minimize the deviation of the Cherenkov hit timing distribution.
The algorithm can cover 97\% of muons passing through the ID, but is not suitable for muon bundles, stopping muons and muons inducing energetic showers.
These are identified as mis-reconstructed muons.

\subsection{\label{sec:shower}Muon-induced showers}
\begin{figure}[h]
    \includegraphics[width=1.0\linewidth]{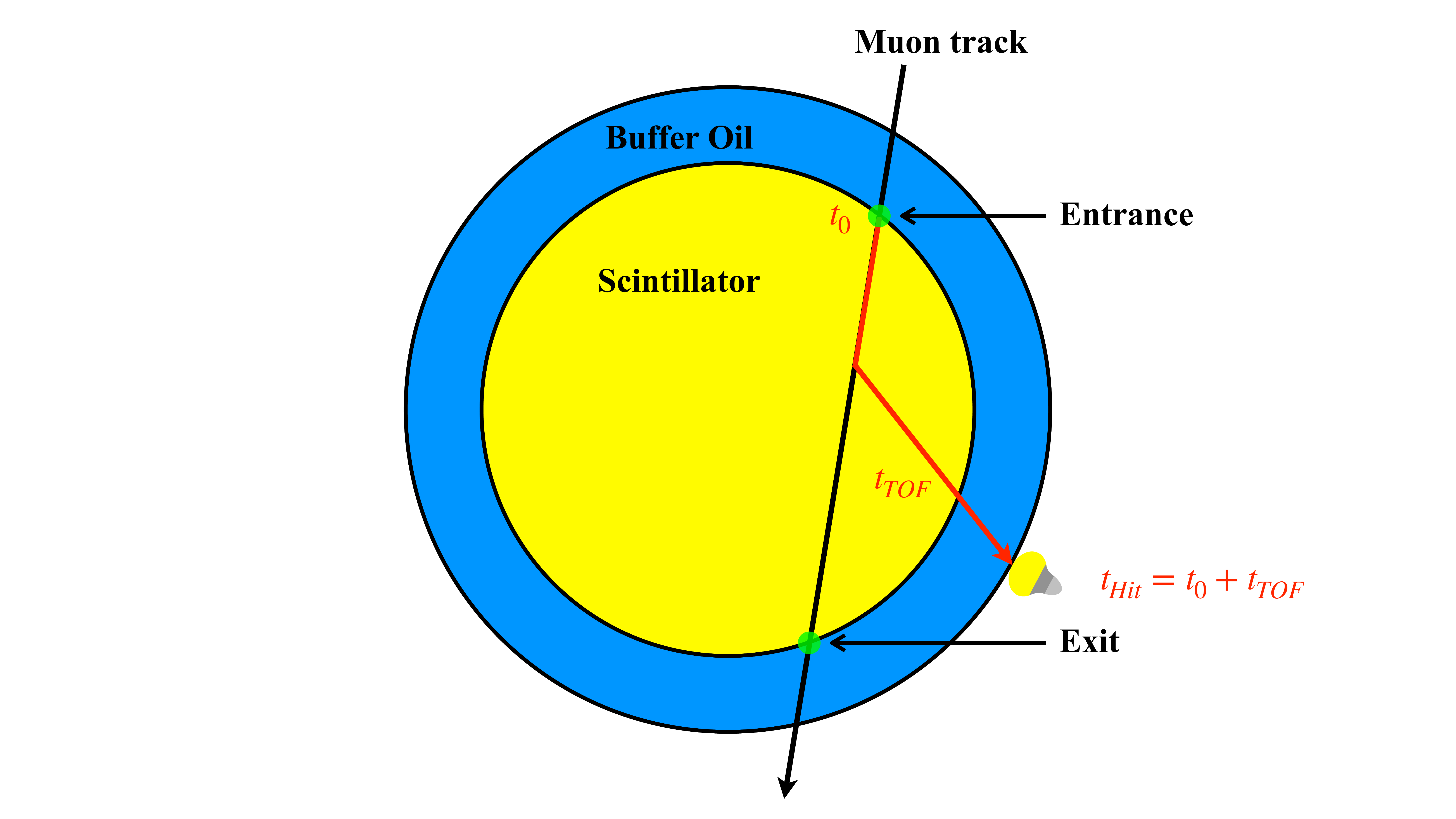}
    \caption{\label{fig:shower_recon}Schematic of the scintillation light path of a muon event. $t_{Hit}$ is the expected hit timing calculated by adding the arrival time and the time-of-flight.}
\end{figure}

\begin{figure}[h]
    \includegraphics[width=1.0\linewidth]{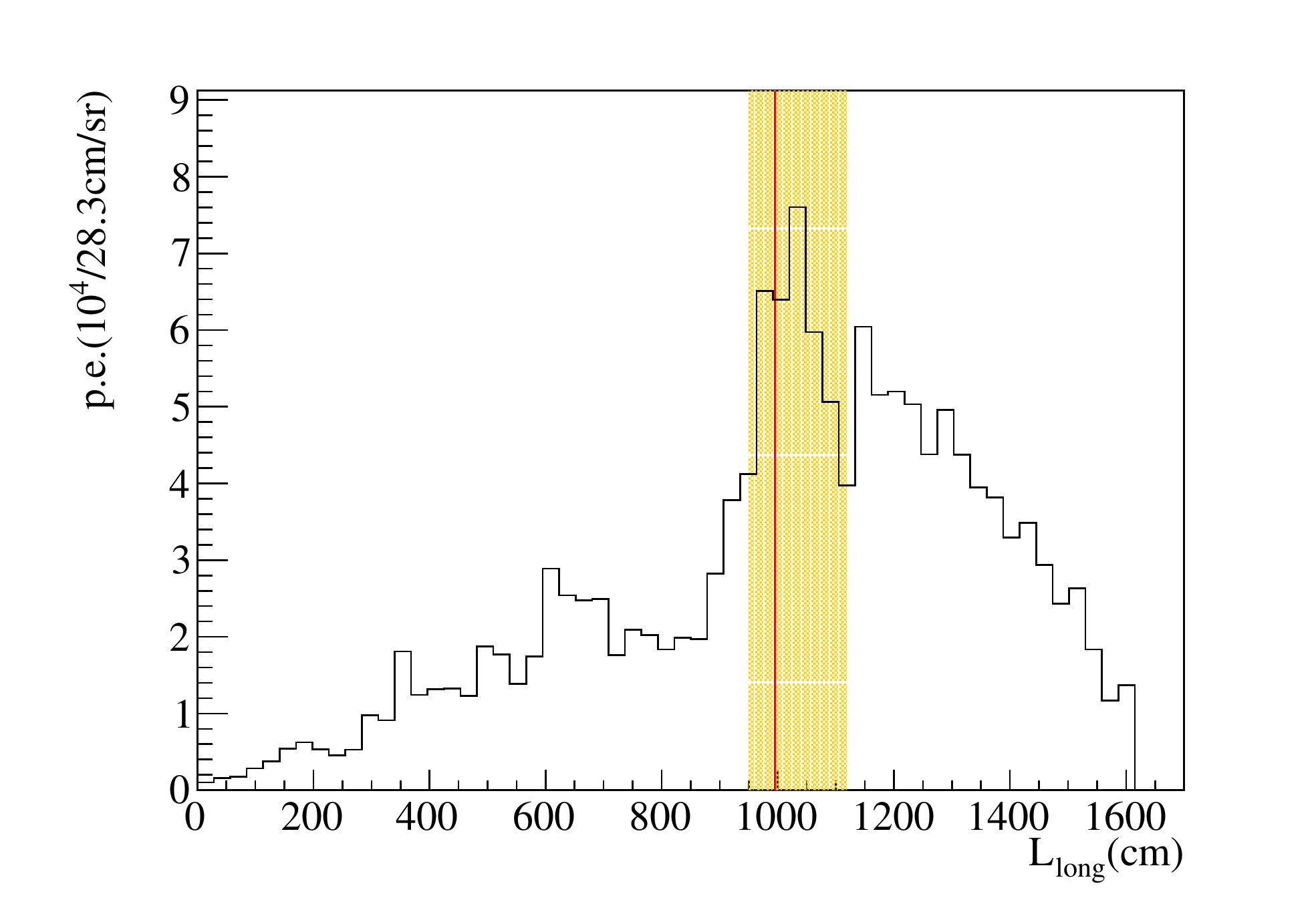}
    \caption{\label{fig:Llong}An example of the reconstructed shower charge along the track of a measured muon.
    The red line ($L_{long}=995$\,cm) is the position of the interaction. As detailed in Sec.~\ref{sec:carbon}, the shower charge, $dQ/dL_{long}=3.96\times10^{5}$\,p.e./cm, is calculated by integrating the deposition in the orange band.}
\end{figure}

Cosmic-ray muons induce particle showers by electromagnetic interactions and by hadronic interactions.
Considering a shower along the muon track, the scintillation light emitted in a direction apart from the Cherenkov angle contributes to the photons which have time-delays relative to the expected hits from Cherenkov angles.
Assuming all photons are produced on the track, time-delay can be translated to shower positions.
The shower charge along the longitudinal distance is extracted from the digital waveform for each PMT.
The expected hit timing ($t_{Hit}$) is calculated by, 
\begin{equation}\label{eq:t_hit}
t_{Hit} = t_{0} + t_{TOF},
\end{equation}
where $t_{0}$ is the time when the muon arrives at the entrance and $t_{TOF}$ is the time-of-flight for a given position on the track (see Fig.~\ref{fig:shower_recon}).
$t_{Hit}$ is compared with the observed waveform, and the position of photon emission is determined by minimizing the difference.
Figure~\ref{fig:Llong} shows an example of the reconstructed shower charge projected onto the longitudinal distance ($L_{long}$) of the muon track.
The peak around $L_{long}\,\sim\,1000$\,cm indicates the reconstructed position of a spallation reaction.

\subsection{\label{sec:NeutronRecon}Muon-induced neutron vertex}
Most of the neutrons observed in KamLAND are initiated by spallation reactions.
They are immediately thermalized and captured by the components of the LS. 
The capture cross sections and the number of target nuclei given in Table~\ref{tab:Ntarget} and Table~\ref{tab:CaptureRatio} show that 99\% of neutrons are captured by $^1$H.
\begin{table}[h]
\caption{\label{tab:Ntarget} Number of target nuclei \,(kton$^{-1}$) in KamLAND-LS and Xe-LS. KamLAND contains 1\,kton of KamLAND-LS and 24\,ton of Xe-LS.}
\begin{ruledtabular}
\begin{tabular}{lcc}
\textrm{Element}&
\textrm{KamLAND-LS}&
\textrm{Xe-LS} \\
\colrule
 Hydrogen & 8.47\,$\times\,10^{31}$ & 8.39\,$\times\,10^{31}$\\
Carbon & 4.30\,$\times\,10^{31}$ & 4.17\,$\times\,10^{31}$\\
Nitrogen & 5\,$\times\,10^{27}$ & 8.31\,$\times\,10^{27}$\\
Oxygen & 5\,$\times\,10^{27}$ & 8.31\,$\times\,10^{27}$\\
Xenon & - & 1.39\,$\times\,10^{29}$\\
\end{tabular}
\end{ruledtabular}
\caption{\label{tab:CaptureRatio}Neutron capture isotopes and their contribution to the total neutron capture in the Xe-LS~\cite{ENSDF}.}
\begin{ruledtabular}
\begin{tabular}{lccc}
\textrm{Nucleus}&
\textrm{$\gamma$\,(MeV)}&
\textrm{Cross section\,(mb)}&
\textrm{(\%)}\\
\colrule
$^1$H & 2.223 & 332.6$\,\pm\,$0.7 & 99.361 \\
$^{12}$C & 4.945 & 3.53$\,\pm\,$0.07 & 0.519\\
$^{13}$C & 8.174 & 1.37$\,\pm\,$0.04 & 0.002 \\
$^{136}$Xe & 4.025 & 238$\,\pm\,$19~\cite{Xecapture} & 0.107\\
$^{134}$Xe & 6.36 & 265.1 & 0.011 \\ 
\end{tabular}
\end{ruledtabular}
\end{table}

The capture $\gamma$-ray events are selected from the data acquired by MoGURA. 
The event vertex reconstruction uses a maximum likelihood estimate. 
We define on-time and off-time windows as follows.
\begin{itemize}
    \item On-time : $-15\,<\,t_i - t_{i_{TOF}}\,<\,15$\,ns,
    \item Off-time : $-100\,<\,t_i - t_{i_{TOF}}\,<\,-15$\,ns or \\
    \hspace*{1.5cm}  $15\,<\,t_i - t_{i_{TOF}}\,<\,100$\,ns,
\end{itemize}
where $t_{i_{TOF}}$ is the time-of-flight between the assumed vertex and $i$-th PMT which detected a hit at $t_{i}$.
The scintillation photon hits are counted in a 200\,ns time window, including multiple hits per PMT.
The contribution of scintillation to the number of hits is estimated by introducing the quantity $N_{S}$ which is defined as the difference of the number of on-time hits ($N_{ON}$) and off-time hits ($N_{OFF}$),
\begin{equation}\label{eq:Ns}
    N_{S} \equiv N_{ON} - N_{OFF} \times \frac{30}{170}.
\end{equation}
The last factor is to compensate for the difference in time-window length.
The vertex reconstruction algorithm loops over all hits. 
$N_{S}$ is repeatedly calculated by shifting the 200\,ns-time window every 20\,ns. 
Finally, the vertex is determined for the position with maximum $N_{S}$.

\begin{figure}[h]
    \includegraphics[width=1.0\linewidth]{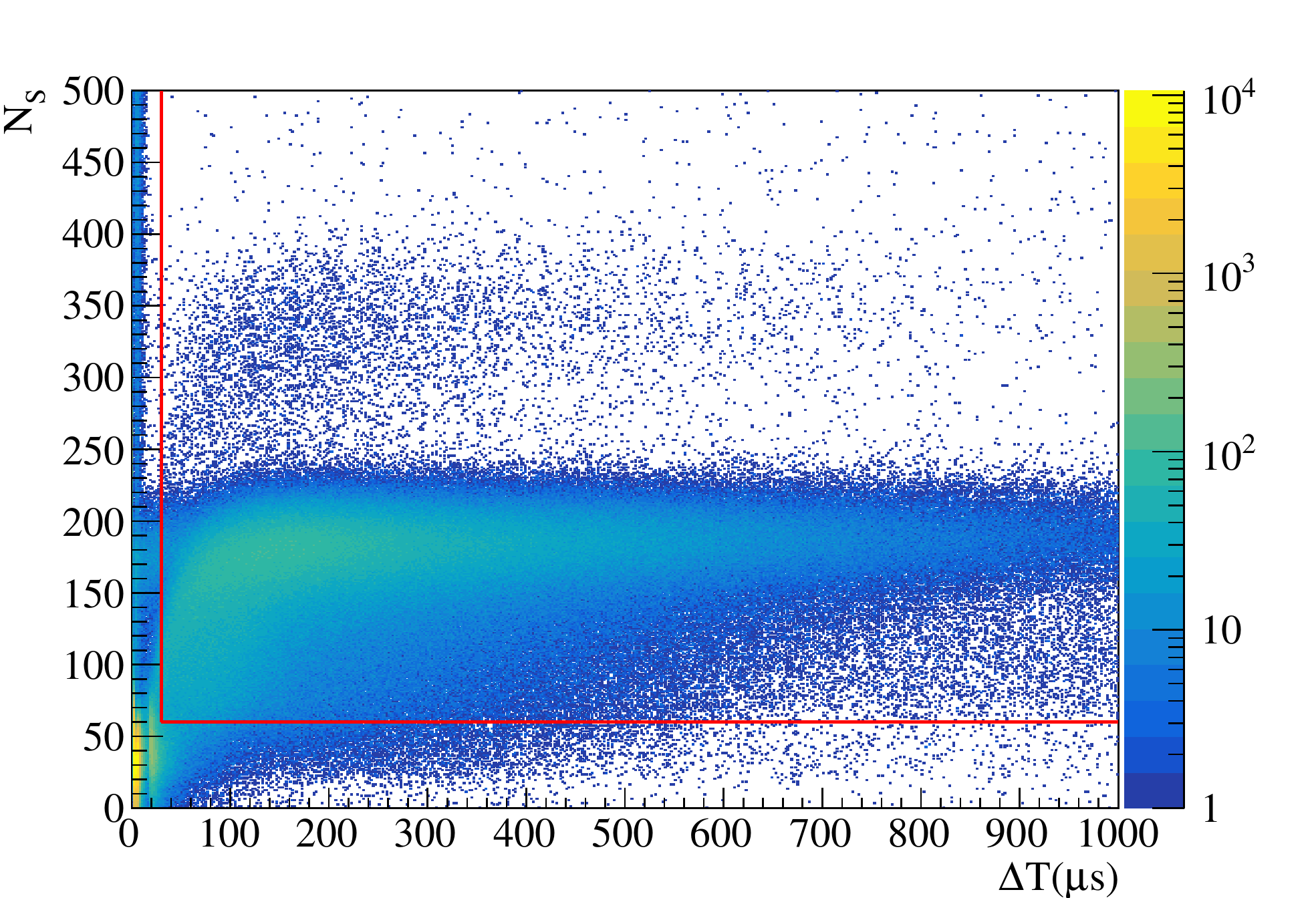}
    \caption{\label{fig:NsdT} $N_{S}$ and $\Delta T$ distribution of neutron capture candidates in a radius less than 5.5\,m. $^1$H$(n, \gamma)^2$H events concentrate around $N_{S}\,\sim\,$180. Noise events are localized around $\Delta T\,<\,30\,\mu$s. The selection criteria for neutron capture events are indicated by the red lines.}
    \includegraphics[width=1.0\linewidth]{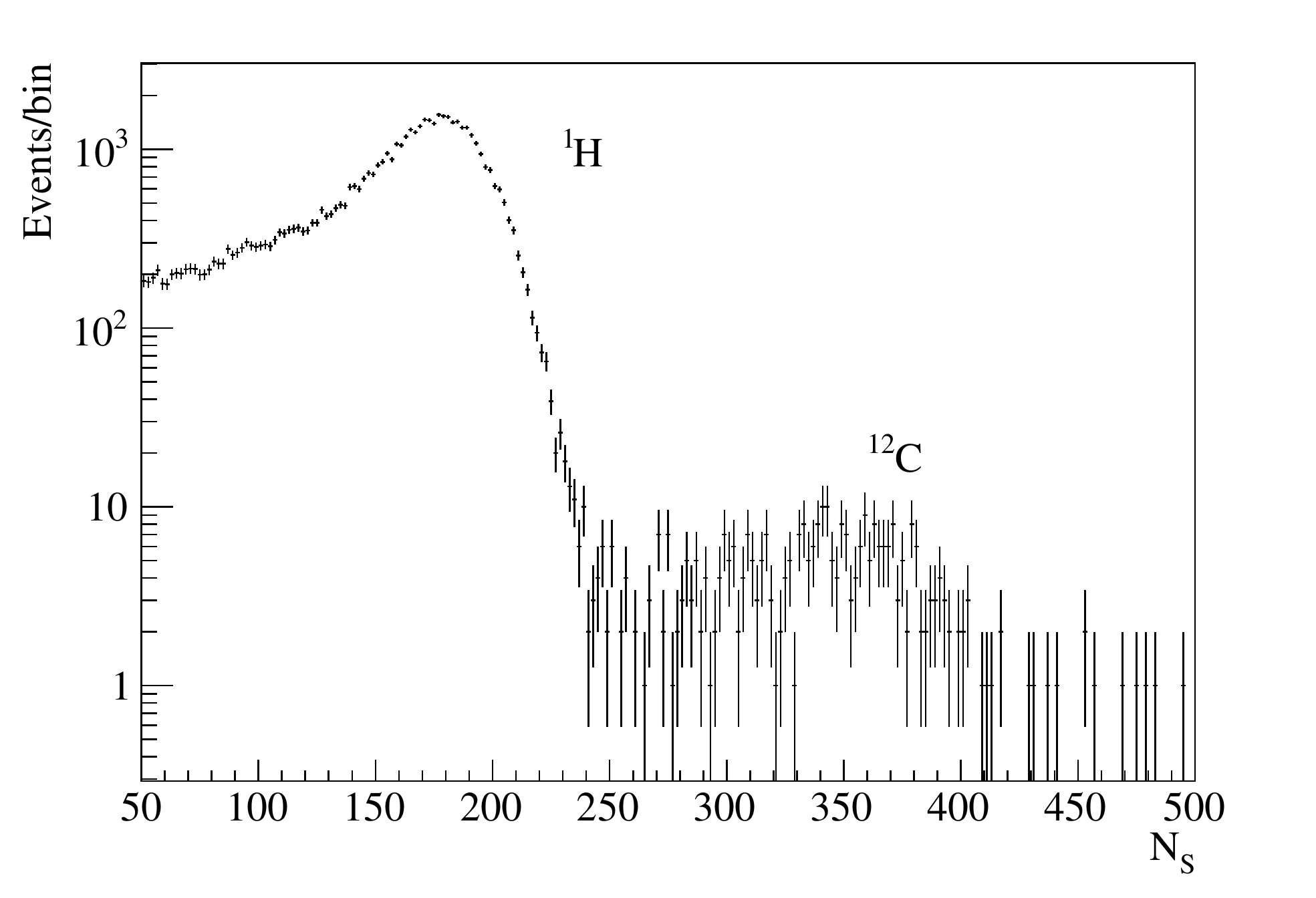}
    \caption{\label{fig:Ns}$N_{S}$ of the neutron capture candidates in the Xe-LS with a radius less than 1.9\,m. The 2.223 (4.945)\,MeV $\gamma$-ray from neutron capture on $^1$H ($^{12}$C) gives a peak around $N_{S}\,\sim\,$180 (350). The 4.025\,MeV $\gamma$-ray from neutron capture on $^{136}$Xe is expected at $N_{S}\,\sim\,$290 but is not visible due to background.}
\end{figure}

Muon-induced neutron captures are extracted from $N_{S}$ and the time difference to the muon ($\Delta T$).
Figure~\ref{fig:NsdT} shows that the neutron capture events on $^1$H are clustering around $N_{S}\,\sim$\,180, while noise events such as after-pulsing and ringing are at $\Delta T\,<\,$30\,$\mu$s.
The neutron capture on $^{12}$C is indicated around $N_{S}\,\sim\,350$ (see Fig.~\ref{fig:Ns}). 
$N_{S}$ gives an estimate of the gamma-ray energy. 

\section{\label{sec:mc}Monte Carlo simulations} 
We study spallation production with two different Monte Carlo simulation tools.
The radioisotopes resulting from muon interaction with the LS are simulated with FLUKA~\cite{FLUKA1, FLUKA2} and subsequently their decay paths are simulated with  Geant4~\cite{Geant4_1, Geant4_2, Geant4_3}.
In this section, we discuss the simulation and uncertainties of spallation isotopes in xenon with FLUKA.
In addition, we discuss the reconstruction of the energy spectrum for each isotope with Geant4.

\subsection{\label{sec:fluka} FLUKA}
FLUKA is a well-studied Monte Carlo simulation tool modeling hadronic interactions. 
We used FLUKA 2011.08.patch to calculate the spallation production yield. 
The physics models that were activated for this purpose, are listed in Table~\ref{tab:fluka}.
The decay of produced radioactive nuclei and isomer production were not activated, since we traced the decay paths with Geant4 as discussed in Sec.~\ref{sec:geant4}.
FLUKA was used together with the heavy ion interaction models, rQMD-2.4~\cite{rQMD} and DPMJET-3~\cite{DPMJET3} by linking the library.
\begin{table*}
\caption{\label{tab:fluka}FLUKA physics processes. Radioactive decay is deactivated in FLUKA since we use Geant4 to calculate it.}
\begin{ruledtabular}
\begin{tabular}{ccc}
\textrm{Card}&
\textrm{Physics}&
\textrm{Status}\\
\colrule
DEFAULTS & A set of physics models & PRECISIO(n)\\
PHOTONUC(lear) & Gamma interactions with nuclei & Activated \\
MUPHOTON & Muon photonuclear interaction & Activated \\
PHYSICS & Emission of light fragments & Activated by COALESCE(nse) \\
PHYSICS & Emission of heavy fragments & Activated by EVAPORAT(ion)\\
PHYSICS & Ion electromagnetic-dissociation & Activated by EM-DISSO(ciation) \\
PHYSICS & Decay and isomer production & Deactivated by RADDECAY\\
\end{tabular}
\end{ruledtabular}
\end{table*}

The reproducibility of muon induced xenon spallation cross sections in FLUKA has not been tested, since there are no measurements of these cross sections yet.
However, the measurement of charged hadron production with a 490\,GeV positive muon beam on gaseous xenon was reported in \cite{E665}. 
DPMJET-3, the MC generator implemented in FLUKA, reproduced the measurement as presented in \cite{DPMJET3} demonstrating the appropriate modeling of the physics processes such as muon-nucleon scattering and hadronization processes.  
FLUKA is widely adopted to estimate production of particles in related fields~\cite{EXO, borexino}.

\begin{figure}[h]
    \includegraphics[width=1.0\linewidth]{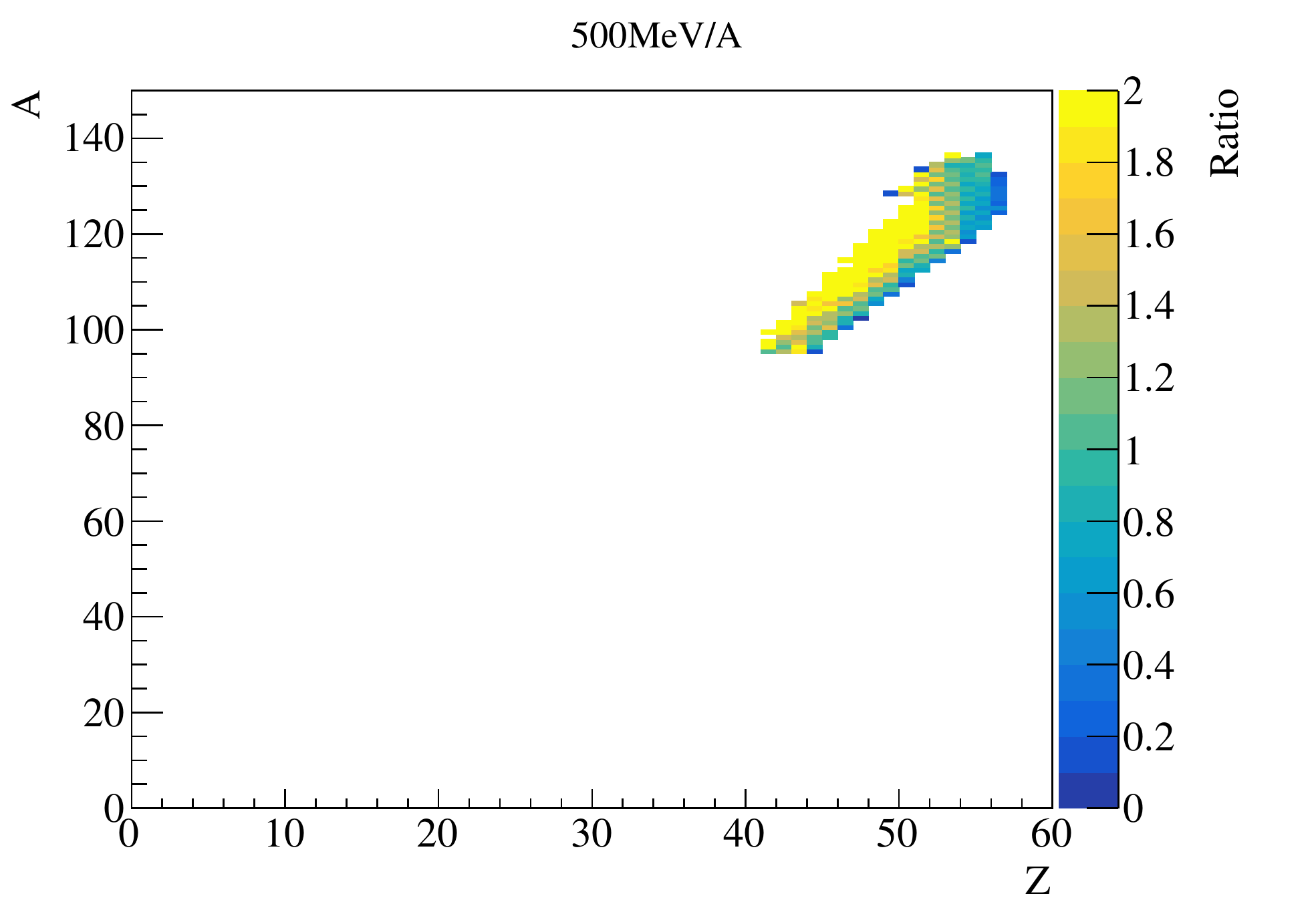}
    \includegraphics[width=1.0\linewidth]{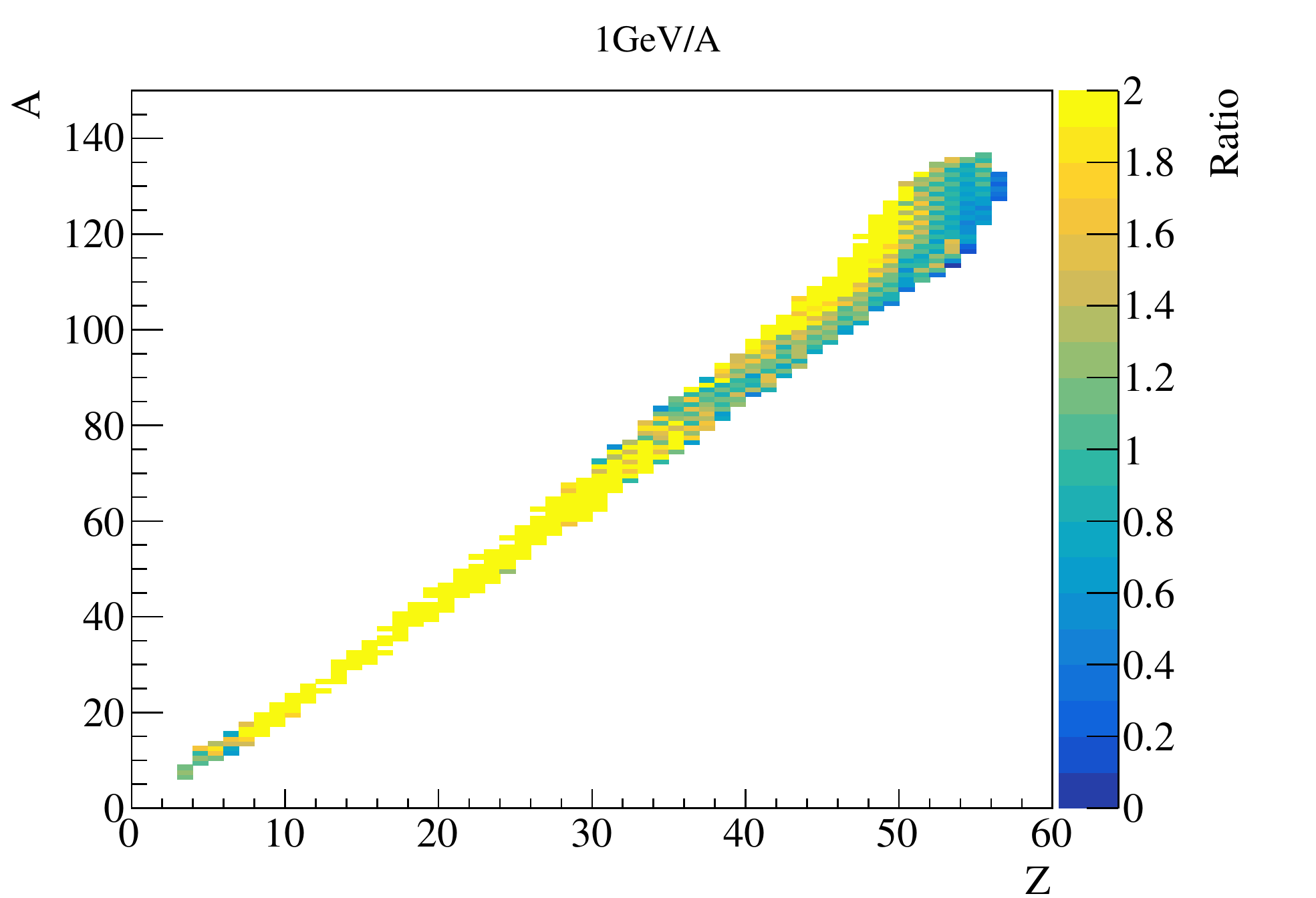}
    \caption{\label{fig:Xsections_ratio}The cross section ratio of the measurements to the FLUKA simulations, for a 500\,MeV/A $^{136}$Xe beam (top) and a 1\,GeV/A $^{136}$Xe beam (bottom). Only $Z>40$ was reported in \cite{500MeV}.}
\end{figure}

To quantitatively estimate the uncertainty on production rates in FLUKA, we compared our simulation with the measurement of residual isotope production from beam experiments.
Production cross sections were measured by irradiating a 1\,cm$^3$-liquid hydrogen target with a $^{136}$Xe beam, where the incident energy-per-nucleon was 500\,MeV~\cite{500MeV} and 1\,GeV~\cite{1GeV}.
The simulation prediction and measured cross sections are compared in Fig.~\ref{fig:Xsections_ratio}.
The deviation for the produced isotopes is large in $7\le Z \le35$ and $40\le Z \le50$, while the simulation shows good agreement in $Z>50$ where the primary production yield is dominant.

\begin{figure}
    \includegraphics[width=1.0\linewidth]{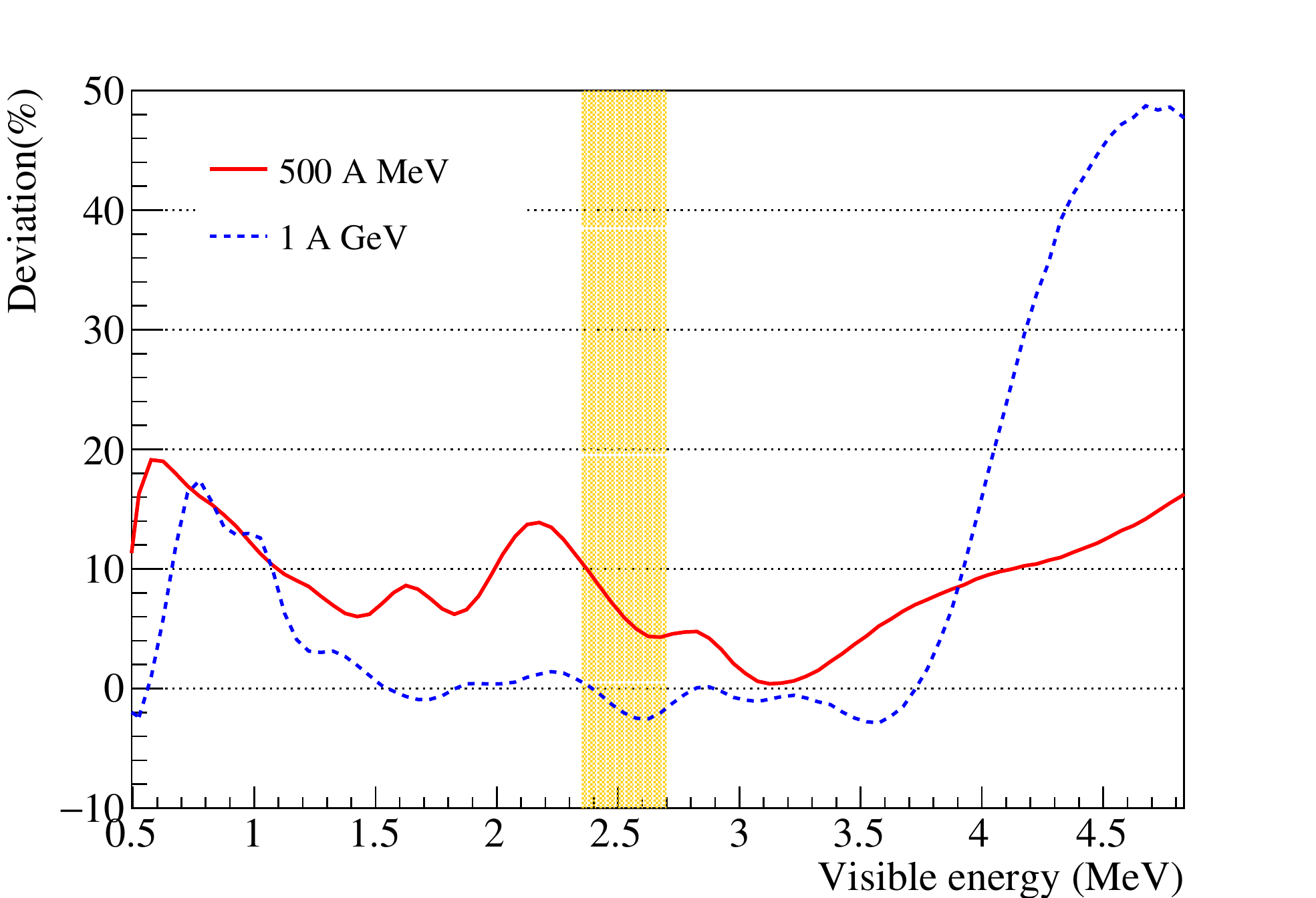}
    \caption{\label{fig:penalty}Deviation of the visible spallation isotope decay energy spectrum compared to FLUKA. The data spectrum is constructed from the measured cross sections of the 500\,MeV/A $^{136}$Xe beam experiment (red solid)~\cite{500MeV} and the 1\,GeV/A $^{136}$Xe beam experiment (blue dashed)~\cite{1GeV}. The ROI is indicated by the band.}
\end{figure}

The decay energy spectrum for each spallation product was determined (see Sec.~\ref{sec:geant4}).
The comparison between data and simulation is shown in Fig.~\ref{fig:penalty} for both the 500\,MeV/A and 1\,GeV/A beams.
Focusing on the region of interest (ROI) for the $0\nu\beta\beta$ decay search ($2.35\,\leq\,E\,\leq\,2.70\,$MeV), the result from the 500\,MeV/A beam shows a larger deviation, which indicates the cross section is larger for data. 
In the spectrum fit for the $0\nu\beta\beta$ decay search~\cite{Zen800}, 
the simulated shape of the xenon spallation spectrum is
allowed to float within the deviation from the 500\,MeV/A beam measurement for a conservative approach.

\begin{figure}
    \includegraphics[width=1.0\linewidth]{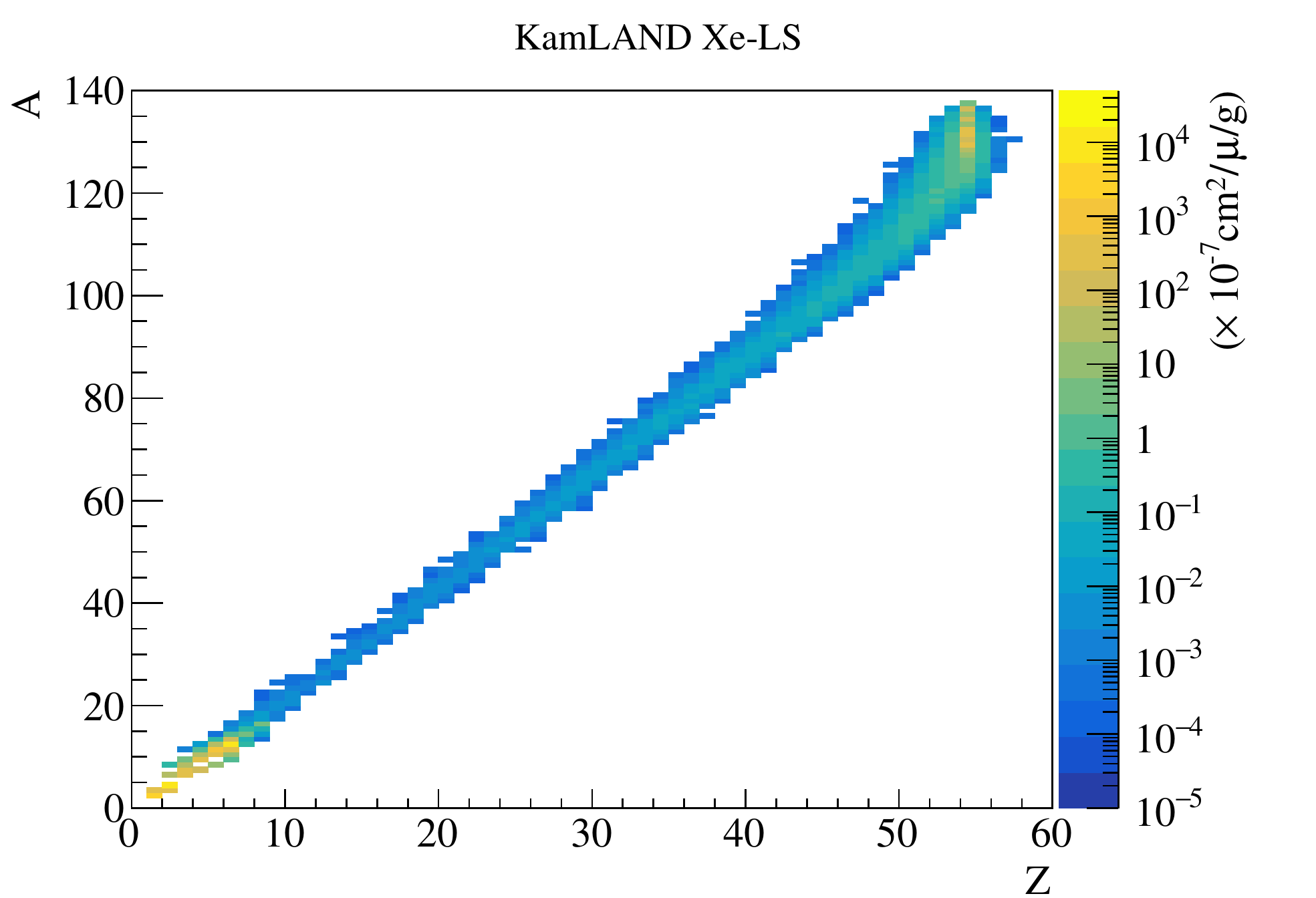}
    \caption{\label{fig:AvsZ} Muon-induced spallation production yields in the Xe-LS from FLUKA simulation.}
\end{figure}

The spallation product yield in the Xe-LS is calculated by emitting 2$\,\times\,10^7$ cosmic-ray muons into a 40\,m-high and 10\,m-radius cylinder. 
The initial muon energy is generated with the MUSIC simulation framework~\cite{MUSIC,MuonatKamLAND,spallation2010,MUSIC2}. 
With a detailed geometric description of Mt. Ikenoyama, MUSIC can calculate muon transportation in rock and output the survival probability. 
We used Gaisser's surface muon flux model~\cite{spallation2010}.
The MUSIC-generated muon angular and energy spectra are sampled and act as input for the FLUKA calculation.
The mean simulated muon energy is $260\,\pm\,1\,$GeV.
The regions 10\,m from the muon entrance and 5\,m from the cylinder exit are removed from analysis in order to avoid boundary effects.
The corresponding detector livetime is 9\,yr as calculated from the total muon track length.
The simulation with the same geometry but made of KamLAND-LS predicts the neutron capture time to be $\tau_n = 207.0\,\pm\,0.3\,\mu$s which is consistent with measurement~\cite{spallation2010}. 
The neutron capture rate is given in Table~\ref{tab:nCapOn1H} and radioisotope production is presented in Fig.~\ref{fig:AvsZ}.
The simulated muon charge ratio is $\mu^{+}/\mu^{-} = 1.3$~\cite{MuMinusflaction}.

\begin{figure}
    \includegraphics[width=1.0\linewidth]{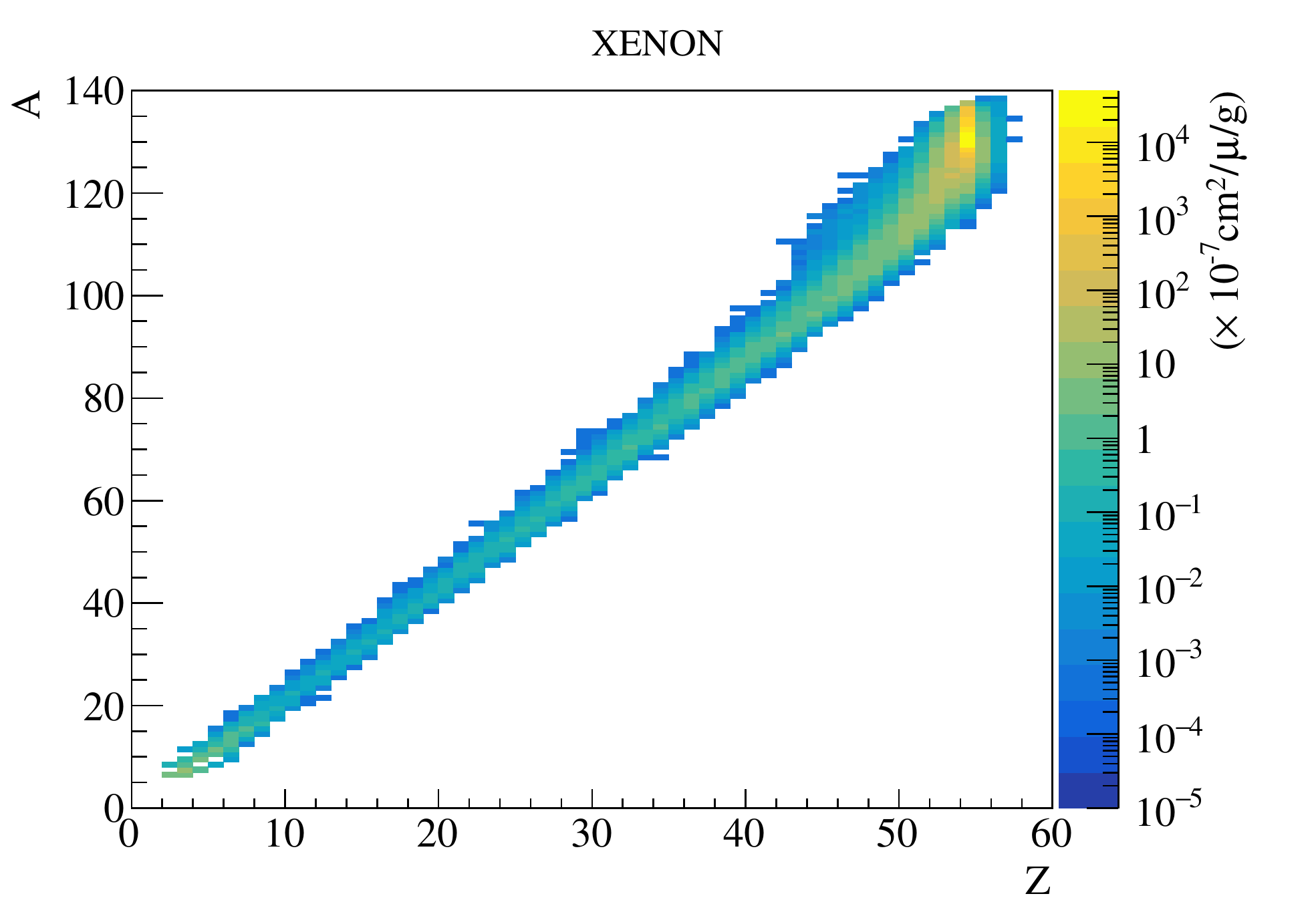}
    \includegraphics[width=1.0\linewidth]{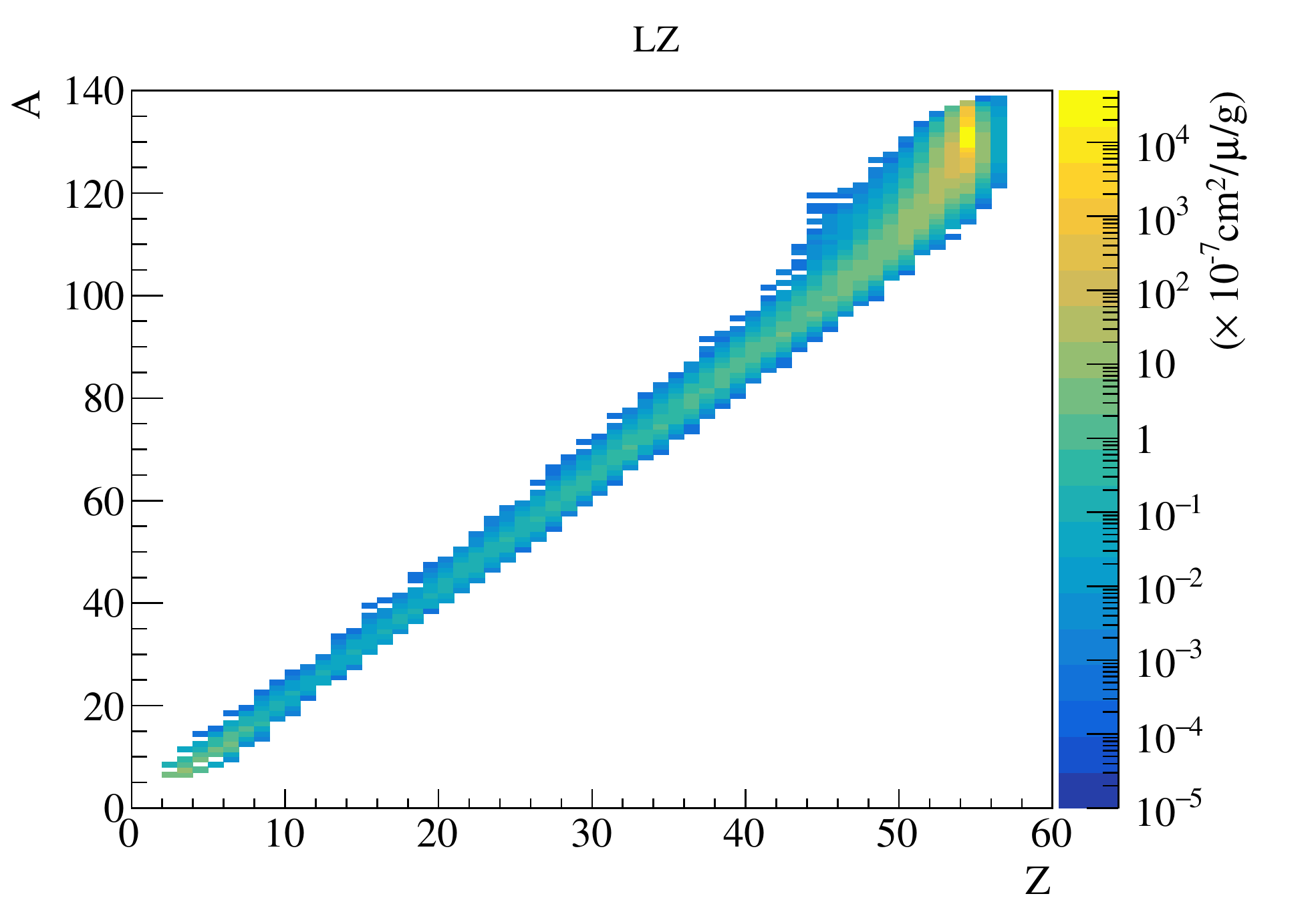}
    \caption{\label{fig:LXe}Muon-induced spallation production yields in the liquid xenon from FLUKA simulations. The initial muon energy is 273\,GeV and 283\,GeV for XENON~\cite{XENON100} (top) and LZ~\cite{LZ} (bottom), respectively. $A\le$4 is omitted.}
\end{figure}

We also provide a prediction of the muon-induced spallation production yield in natural liquid xenon experiments (see Fig~\ref{fig:LXe}).
The simulation configuration is similar to Xe-LS, but 
only muons of a fixed energy are injected into a cylinder made of liquid xenon.
We note that FLUKA 2021.2.7 was used for the liquid xenon simulation in Fig.~\ref{fig:LXe}.

\subsection{\label{sec:geant4} Geant4}
The xenon spallation products include long-lived isotopes.
Unlike carbon, xenon can provide various unstable isotopes because of its large mass number, so sequential decays have to be taken into account for a comprehensive understanding of the backgrounds.

\begin{figure}
    \includegraphics[width=1.0\linewidth]{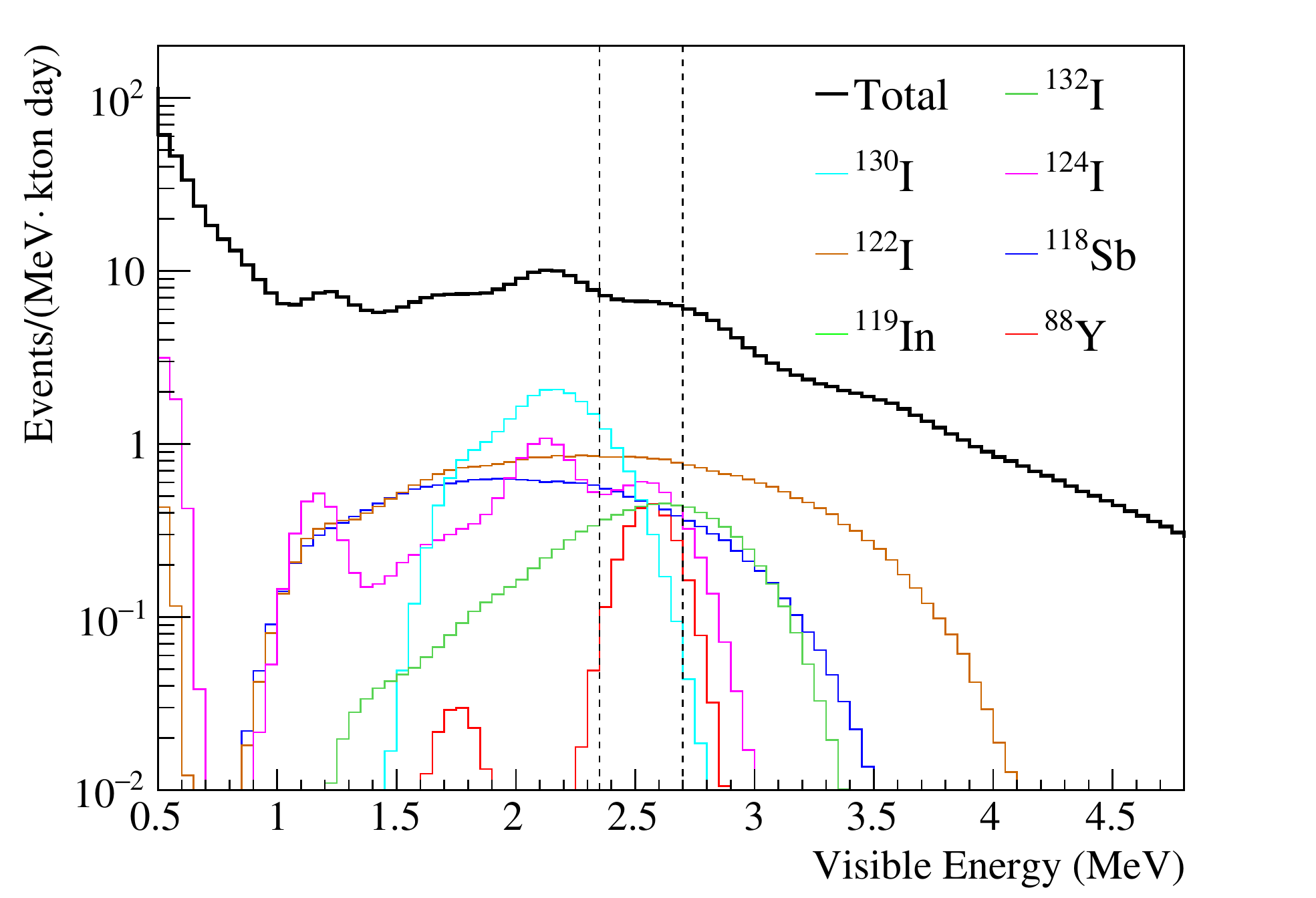}
    \caption{\label{fig:EnergySpectra} Simulated visible energy spectra of spallation isotopes in Xe-LS. The colored histograms show the contribution from a part of the spallation daughters which have large impacts in the ROI. The black histogram is the total of all spallation isotopes. The $0\nu\beta\beta$ decay search ROI is indicated by black dashed lines. }
\end{figure}

We use Geant4 to estimate the radioactive decay times and to reconstruct the energy spectra.
In this calculation, Geant4 version geant4.10.6.p01 is used with version G4ENSDFSTATE2.2 for the Evaluated Nuclear Structure Data File (ENSDF)~\cite{ENSDF}.
We confirmed the $Q$-values, half-lifes and branching ratios for more than a thousand isotopes by comparing the Geant4 version with the online database version of ENSDF.
All sequential decay products and their energy deposits were recorded and the visible energy spectrum for each isotope was simulated including the detector energy resolution and quenching effect.
For each of the spallation products listed by FLUKA, 10$^6$ events were generated to trace the decay path. 
The energy spectra of some xenon isotopes are shown in Fig.~\ref{fig:EnergySpectra}.
The simulation is executed without defining a detector geometry.

\section{\label{sec:data}Spallation production}
This section discusses neutron capture rates and spallation productions of carbon and xenon.
Since the statistics in Xe-LS is too low, the carbon spallation is estimated from KamLAND-LS data. 
We used FLUKA predictions for the selection efficiencies.
The xenon spallation is estimated directly from the Xe-LS data, but the decomposition of individual isotopes is not possible due to the limited statistics.
The uncertainties on the relative contributions are estimated based on the beam data (see Sec.~\ref{sec:fluka}).

\subsection{\label{sec:NeutronCapture}Neutron capture on $^{1}$H and $^{12}$C}
\begin{figure}[h]
    \includegraphics[width=1.\linewidth]{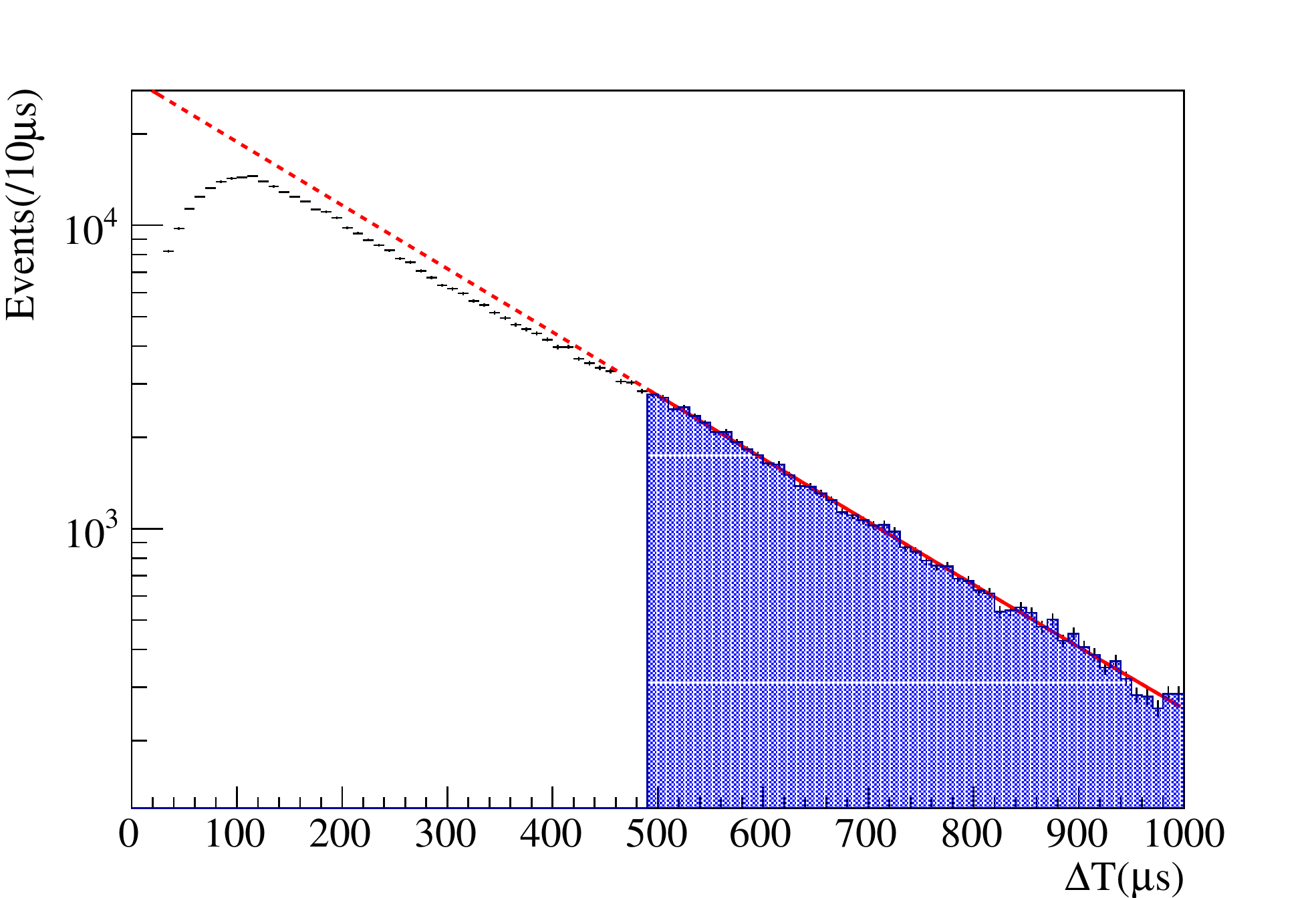}
    \caption{\label{fig:dT} Time difference between the cosmic-ray muon and neutron capture in the KamLAND-LS ($2.5<r<4.5\,$m with the tube cut). The red line shows the exponential fit constrained to 207.5$\pm2.8\,\mu$s. Fit range is illustrated by the shaded region. The detection efficiency is lower at small $\Delta T$ due to PMT baseline overshoot.}
\end{figure}

\begin{figure}
    \includegraphics[width=1.\linewidth]{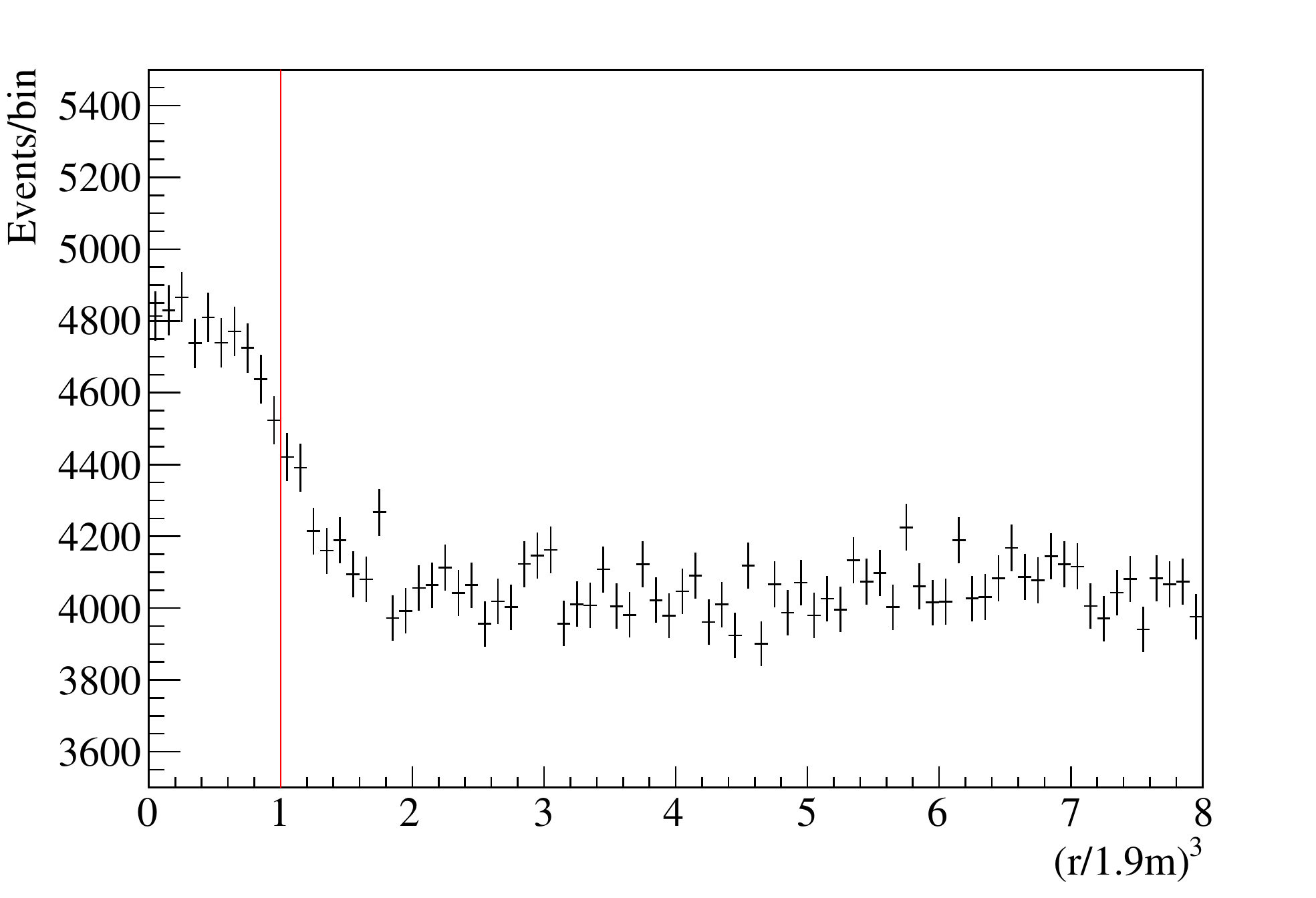}
    \caption{\label{fig:r3}Spatial distribution of neutron capture events. $r$ is the distance from the detector center. The boundary of Xe-LS and KamLAND-LS is indicated by the red line. The vertex resolution and droplet-like shape of the inner balloon cause the deviation from a step function.}
\end{figure}

The number of neutron captures ($N_n$) is obtained by fitting to the data with a function of the mean capture time ($\tau_n$):
\begin{equation}\label{eq:dTfit}
    \mathcal{R}(t) = \frac{N_{n}}{\tau_{n}} e^{-(t-t_{\mu})/\tau_n} + const,
\end{equation}
where $\tau_{n}$ is constrained to $207.5\,\pm\,2.8\,\mu$s~\cite{spallation2010} with a Gaussian penalty $\chi^2$-term. 
The fit range is set to 490$\,\leq\,\Delta T\,<\,1000\,\mu$s ($\Delta T \equiv t-t_{\mu}$) to meet the constraint and to avoid after-pulsing and PMT baseline overshoot effects.
Figure~\ref{fig:dT} shows the time correlation between a cosmic-ray muon and subsequent neutron captures, where the neutron capture events are selected with $30\,<\Delta T<\,1000\,\mu$s and $N_{S}\,>\,60$ in the KamLAND-LS outside of the IB ($2.5<r<4.5$\,m). 
In addition to these selections, we apply the tube cut shown in Fig.~\ref{fig:kamland}.
The neutron capture rate in the KamLAND-LS is found to be: 
\begin{eqnarray}\label{eq:Nrate}
\mathcal{R}_{n} = \frac{N_{n}}{\rho VT} = 
3781\,\pm\,28\,\,\rm{(kton\,day)^{-1}},
\end{eqnarray}
where $\rho$\,=\,0.780\,g/cm$^{3}$ and $V$\,=\,301.4\,m${^3}$ are the density and volume of the LS, 
respectively, and $T$ = 711\,day is the detector livetime. 
More generally, the neutron capture rate (and all other rates provided below) can be converted to units of ($\mu^{-1}$\,g$^{-1}$\,cm$^{2}$), using the rate of muons passing through the KamLAND-LS, $\mathcal{R}_{\mu} = 0.198\,\pm\,0.014\,$Hz, and the mean track length, $L_{\mu} = 874\,\pm\,13\,$cm~\cite{spallation2010},
\begin{eqnarray}\label{eq:Nrate2}
\mathcal{R}_{n} &=& \frac{N_{n}}{\mathcal{R}_{\mu} T \rho L_{\mu}} \notag \\
&=& (2.91\,\pm\,0.02) \times10^{-4}
\rm{(\mu^{-1}\,g^{-1}\,cm^{2})}.
\end{eqnarray}
Similarly, the neutron capture rate in the Xe-LS is obtained by selecting $r<1.9$\,m with the hot spot veto.
The results are summarized in Table~\ref{tab:nCapOn1H}, together with the FLUKA predictions.
\begin{table}[h]
\caption{\label{tab:nCapOn1H}Neutron capture rate in the LS with statistical uncertainty. The measured values have a 7.8\,\% systematic uncertainty derived from the fiducial volume, while FLUKA has a 7.2\,\% systematic error from the estimates of $\mathcal{R}_{\mu}$ and $L_{\mu}$.}
\begin{ruledtabular}
\begin{tabular}{ll}
\textrm{}&
\textrm{KamLAND-LS\,(kton\,day)$^{-1}$}\\
\colrule
Data & $3781\pm28\rm{(stat.)}\pm295\rm{(syst.)}$ \\
FLUKA & $4046\pm1\rm{(stat.)}\pm292\rm{(syst.)}$  \\
\hline\hline
\textrm{}&
\textrm{Xe-LS\,(kton\,day)$^{-1}$}\\
\colrule
Data & $4347\pm98\rm{(stat.)}\pm339\rm{(syst.)}$ \\
FLUKA & $4647\pm1\rm{(stat.)}\pm336\rm{(syst.)}$ \\
\end{tabular}
\end{ruledtabular}
\end{table}
As shown in Fig.~\ref{fig:r3}, the difference in capture rate between Xe-LS and KamLAND-LS appears to follow a $r^{3}$ distribution, which is in agreement with the estimation of FLUKA.

\subsection{\label{sec:carbon} Carbon spallation}
Cosmic-ray muons produce radioactive isotopes and neutrons in the detector.
A portion of the carbon spallation backgrounds are identified by a three-fold decay coincidence ($n$-$tag$) of a LS muon (1), muon-induced neutron captures (2) and a $\beta$ decay (3) with the following cuts:
\begin{itemize}
    \item[(1)] $Q_{ID}>40000$\,p.e.,
    \item[(2)] $N_{S}>60$ and $30<\Delta T_{\mu-n}<1000\,\mu$s,
    \item[(3)] $\Delta R_{min}<1.6$\,m.
\end{itemize}
Here, $\Delta T_{\mu-n}$ is the time difference between a cosmic-ray muon and neutron capture, while $\Delta R_{min}$ is the distance between the $\beta$ decay and the nearest neutron capture. 
On the other hand, the time correlation between muon and isotope decay ($t-t_{\mu}$) is used in Eq.~(\ref{eq:Cfit}).

n-tag can discriminate neutron emitting interactions, while non-neutron emitting interactions and the $\beta$ decays isolated from neutrons ($\Delta R_{min}\ge1.6$\,m) are covered by a shower likelihood method ($shower$-$tag$).
The likelihood is constructed as a function of the reconstructed shower charge ($Q_{shower}(L_{long})$) and transverse distance from the muon track ($L_{trans}$) based on the position of the isotope decay candidate events.
The probability density function (PDF) of charge along the muon track ($dQ_{shower}/dL_{long}$) is provided by the shower reconstruction as described in Sec.~\ref{sec:shower}.
A 1.7\,m window is opened around the maximum of $dQ_{shower}/dL_{long}$ and the $Q_{shower}(L_{long})$ is defined as the integral of $dQ_{shower}/dL_{long}$ in the window.
The likelihood function for carbon spallation products was created based on $^{12}$B $\beta^{-}$ decay events, which are not selected by n-tag.
$^{12}$B is separated by requiring a visible energy of more than 6\,MeV and the time difference from a muon to be less than 150\,ms. 
The function for accidental events is constructed assuming uniform distribution in time.

The selection efficiency of the shower likelihood method ($\epsilon_{i-shower}$) depends on the isotope, which in turn can be derived from the interaction difference of the primary particles. 
For example, a neutron is the most probable particle to generate $^{12}$B and it prefers long $L_{trans}$.
Pions are the leading particles in $^{10}$C creation and they distribute closer to the muon track.
That causes differences in the selection efficiencies between isotopes for given $L_{trans}$.
The likelihood PDFs were created with $^{12}$B events because of high statistics, but the fraction of $\mu^{-}$ capture process causes isotope dependence in the PDFs.
To suppress the influence, we refer to the efficiency of $^{8}$Li events obtained by (b) and (c), described later.
The shower-tag is applied for $^{10}$C, $^{6}$He and $^{11}$Be, and their efficiencies are estimated as the discrepancies from $^{8}$Li, calculated with FLUKA.

We use a binned maximum likelihood to analyze our data. 
It is a function of production yield of the $i$-th isotope $\mathcal{N}_{i}$, lifetime $\tau_{i}$ and decay energy: 
\begin{equation}\label{eq:Cfit}
    \frac{d\mathcal{N}}{dt} = \sum_{i} \frac{\mathcal{N}_{i}}{\tau_i} e^{-(t-t_\mu)/\tau_{i}} + const.
\end{equation}
The observed data is filled into log scale time bins and simultaneously fit with energy bins.
Prior to this fit, the $\beta$ decay candidates are categorized into three groups, in order to reduce the statistical uncertainties due to accidental background subtractions:
\begin{itemize}
    \item [(a)] Selected by n-tag ($\mathcal{N}_{i-{n-tag}}$).
    \item [(b)] Not selected by n-tag ($\mathcal{N}_{i-isolated}$).
    \item [(c)] Not selected by n-tag and selected by shower-tag ($\mathcal{N}_{i-shower}$).
\end{itemize}
Note that (b) is the complement of (a), and (c) is a subset of (b).
Equation~(\ref{eq:Cfit}) is used for each category and the production rates obtained by,
\begin{itemize}
    \item[(A)] $\mathcal{N}_i = \mathcal{N}_{i-n-tag} + \mathcal{N}_{i-isolated}$ or
    \item[(B)] $\mathcal{N}_i = \mathcal{N}_{i-n-tag} +\mathcal{N}_{i-shower}/\epsilon_{i-shower}$.
\end{itemize}
Background discrimination is more difficult for isotopes with a long lifetime, such as $^{10}$C, $^{6}$He and $^{11}$Be.
That is why the production rates of those three are estimated using (B), while the others are calculated with (A).
Neutron emitters ($^{8}$He and $^{9}$Li) are estimated first, and short-lived isotopes ($^{12}$B and $^{12}$N) are second, because they have better separation.
We set constraints on these four isotopes in the fit for the other isotopes.

\begin{figure}[h]
    \includegraphics[width=1.0\linewidth]{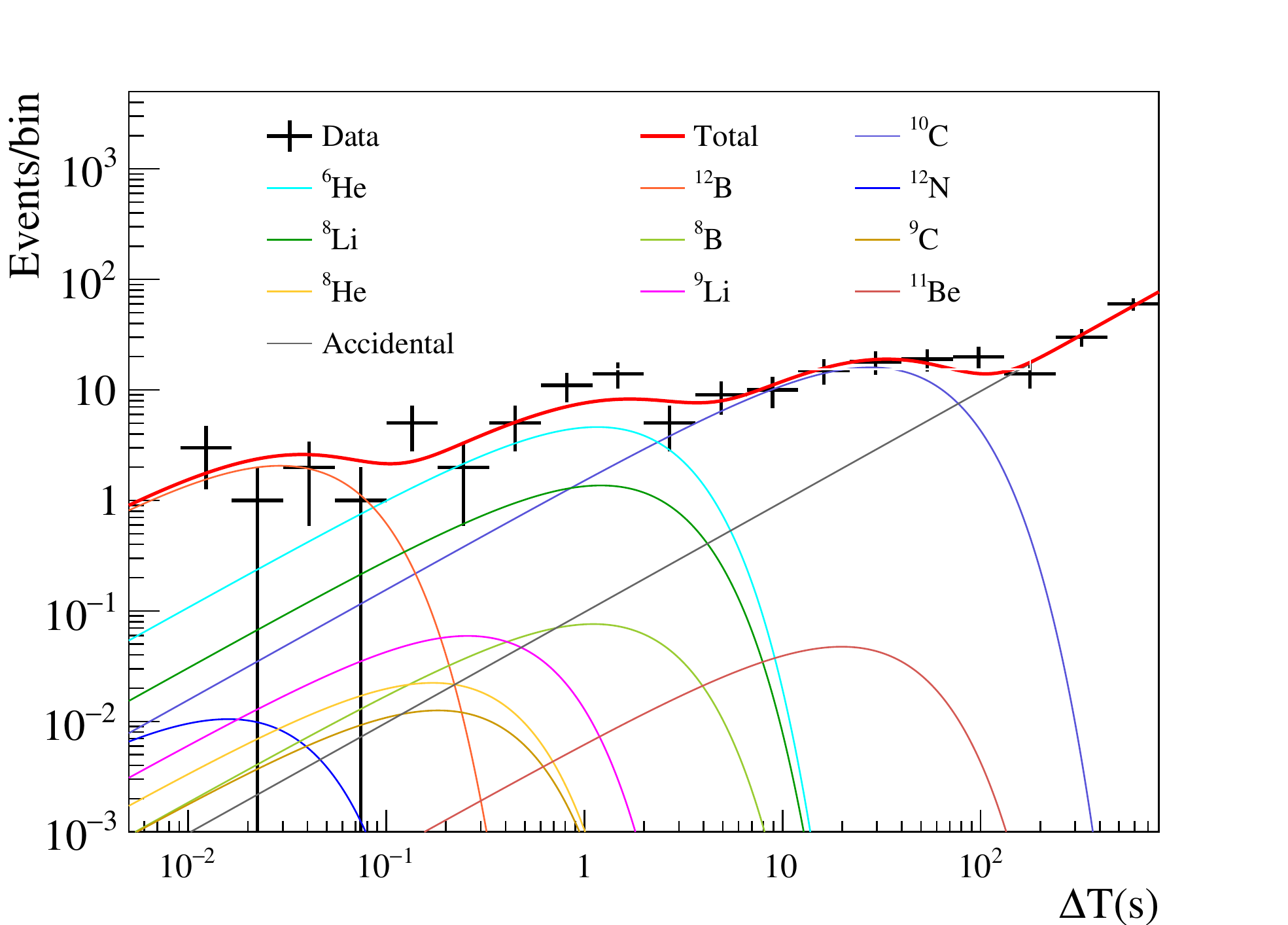}
    \caption{\label{fig:shortfit} Time difference between the cosmic-ray muon and the spallation isotope's $\beta$ decay selected by n-tag. The energy bin is 2$\,<\,E\,<\,3\,$MeV and the $\beta$ decay candidates in $2.5\,<\,r\,<\,4.0$\,m with the tube cut are shown.}
\end{figure}

Figure~\ref{fig:shortfit} shows one example of the $\Delta T$ fit result.
Total production rate (kton\,day)$^{-1}$ of $i$-th isotope is given as,
\begin{eqnarray}\label{eq:Crate}
\mathcal{R}_{i} = \frac{\mathcal{N}_{i}}{\rho VT},
\end{eqnarray}
where the definitions of $\rho, V$ and $T$ are the same as in  Eq.~(\ref{eq:Nrate}).
The isotope production yield in the KamLAND-LS is summarized in Table~\ref{tab:CarbonSummary}. 
The yield in the Xe-LS is also calculated with FLUKA and the differences in carbon spallation products are a few percent, while the neutron capture rate is about 15\% higher in the Xe-LS.

\begin{table*}
\caption{\label{tab:CarbonSummary} Summary of carbon spallation production rate and neutron capture rate in the KamLAND-LS. A unit conversion factor is provided as (kton\,day)$^{-1}$ = 7.69\,$\times\,10^{-8}$\,($\mu^{-1}\,$g$^{-1}$\,cm$^{2}$). 
Spallation reactions were simulated for a KamLAND-LS cylinder and a Xe-LS cylinder, and the results are compared in the last column.
For the precise estimation of $^{11}$C $\beta^{+}$ decay, we require large statistics and the results from KamLAND is presented \cite{7Be}.}
\begin{ruledtabular}
\begin{tabular}{lcc|cccc|c}
\multicolumn{3}{c|}{} & \multicolumn{4}{c|}{Production rate\,(kton\,day)$^{-1}$} & \multicolumn{1}{c}{} \\
\textrm{}&
\textrm{$\tau_{1/2}$}&
\textrm{$Q$\,(MeV)}&
\textrm{$\mathcal{R}_{n-tag}$}&
\textrm{$\mathcal{R}_{isolated}$}&
\textrm{$\mathcal{R}^{Data}_{Total}$}&
\textrm{$\mathcal{R}_{KamLAND-LS}^{FLUKA}$}&
\textrm{$\mathcal{R}_{Xe-LS}^{FLUKA}$/$\mathcal{R}_{KamLAND-LS}^{FLUKA}$}\\
\colrule
$^{8}$He & 119.1\,ms & 10.7\,($\beta^{-}$) & 0.5\,$^{+1.1}_{-0.5}$ & 0.3\,$^{+0.5}_{-0.3}$ & 0.8\,$^{+1.2}_{-0.5}$& 0.55$\pm$0.04 & 0.96$\pm$0.03\\
$^{9}$Li & 178.3\,ms & 13.6\,($\beta^{-}$) & 2.5\,$^{+0.3}_{-0.3}$ & 0.1\,$^{+0.2}_{-0.1}$ & 2.7\,$^{+0.3}_{-0.4}$& 4.9$\pm$0.4 & 1.00$\pm$0.01\\ 
$^{12}$B & 20.2\,ms & 13.4\,($\beta^{-}$) & 43\,$^{+3}_{-2}$ & 14.4\,$^{+0.6}_{-0.08}$ & 58\,$^{+3}_{-2}$& 42$\pm$3 & 1.013$\pm$0.003\\
$^{12}$N & 11.0\,ms & 17.3\,($\beta^{+}$) & 0.8\,$^{+0.9}_{-0.7}$ & 0.07\,$^{+0.2}_{-0.06}$ & 0.9\,$^{+0.9}_{-0.7}$& 0.74$\pm$0.06 & 1.02$\pm$0.03\\
$^{8}$Li & 839.9\,ms & 16.0\,($\beta^{-}$) & 21\,$^{+4}_{-3}$ & 0.9\,$^{+1.6}_{-0.9}$ & 22\,$^{+4}_{-4}$& 47$\pm$3 & 1.021$\pm$0.004\\ 
$^{8}$B & 770\,ms & 18.0\,($\beta^{+}$) & 3\,$^{+4}_{-3}$ & 0.8\,$^{+1.0}_{-0.6}$ & 4\,$^{+4}_{-3}$& 11.0$\pm$0.8 & 1.024$\pm$0.008\\ 
$^{9}$C & 126.5\,ms & 16.5\,($\beta^{+}$) & 1\,$^{+2}_{-1}$ & 0.2\,$^{+0.5}_{-0.2}$ & 2\,$^{+2}_{-1}$& 1.5$\pm$0.1 & 1.02$\pm$0.02\\ 
$^{11}$Be & 13.8\,s & 11.5\,($\beta^{-}$) & 1.2\,$^{+0.3}_{-0.2}$ & 0\,$^{+0.06}_{-0}$ & 1.2\,$^{+0.3}_{-0.2}$& 1.09\,$\pm$0.08 & 1.02$\pm$0.02\\
$^{10}$C & 19.29\,s & 3.65\,($\beta^{+}$) & 19\,$^{+2}_{-2}$ & 0.5\,$^{+0.6}_{-0.5}$ & 19\,$^{+2}_{-2}$& 23$\pm$2 & 1.029$\pm$0.005\\ 
$^{6}$He & 806.7\,ms & 3.51\,($\beta^{-}$) & 10\,$^{+1}_{-1}$ & 1\,$^{+1}_{-1}$ & 11\,$^{+2}_{-1}$& 28$\pm$2 & 1.022$\pm$0.005\\ 
$^{11}$C & 1221.8\,s & 1.98\,($\beta^{+}$) & - & - & 973$\pm$10~\cite{7Be} & 679$\pm$49 & 1.012$\pm$0.001\\ 
$n$ & 207.5\,$\mu$s & 2.223(cap.$\gamma$) & - & - & 3781$\pm$296 & 4046$\pm$292 & 1.1485$\pm$0.0004 \\ 
\end{tabular}
\end{ruledtabular}
\end{table*}

\subsection{\label{sec:xenon} Xenon spallation}
This section discusses xenon spallation with the xenon isotope composition given in Table~\ref{tab:xenon}.
$^{137}$Xe will be addressed in Sec.~\ref{sec:ncaptureXe136}.
The spallation radioisotopes can be divided into two categories: isotopes directly created by spallation and their daughter isotopes.
The former is given by the FLUKA simulation and Geant4 is used for the latter. 
We assume that the abundance of daughter nuclei are in equilibrium, so that the production yield of the $n$-th daughter particle in a decay chain is given by,
\begin{equation}\label{eq:chain}
    \mathcal{N}_{n}(t,\mathcal{N}_0,r) = \frac{r}{\lambda_n} + \frac{1}{\lambda_n}\sum_{i=1}^{n}(\mathcal{N}_0-\frac{r}{\lambda_i})\prod_{j\,\neq\,i}^{n}(\frac{\lambda_j}{\lambda_j - \lambda_i})  e^{-\lambda_i t}.
\end{equation}
$\mathcal{N}_0$ is the initial abundance of the parent isotope, $r$ is the continuous production rate of the parent and $\lambda_n$ is the decay constant for the $n$-th isotope.
Production yields of daughters are calculated using $RadioactiveDecay$ in Geant4.
The equilibrium of a parent isotope can be approximated as a 2000 days exposure of cosmic-ray muons to Xe-LS.

The dominant backgrounds due to xenon spallation in the ROI are enumerated in Table~\ref{tab:criticalLonglived}.
Many isotopes have lifetimes of a few days or longer. 
A simple delayed coincidence method for rejection is therefore not practical considering the detector livetime. 
\begin{table}[h]
\caption{\label{tab:criticalLonglived}Simulated production rate of dominant isotopes in 2.35$\,\leq\,E\,\leq\,$2.70\,MeV in Xe-LS.}
\begin{ruledtabular}
\begin{tabular}{lcc|cc}
\multicolumn{1}{c}{} & \multicolumn{2}{c|}{} & \multicolumn{2}{c}{(kton\,day)$^{-1}$} \\
\textrm{}&
\textrm{$\tau_{1/2}$\,(s)}&
\textrm{$Q$\,(MeV)}&
\textrm{ROI}&
\textrm{Total}\\
\colrule
$^{88}$Y & 9.212$\,\times\,10^{6}$ & 3.62\,(EC/$\beta^{+}\gamma$) & 0.110 & 0.136  \\
$^{90m1}$Zr & 8.092$\,\times\,10^{-1}$ & 2.31\,(IT) & 0.012 & 0.093 \\
$^{90}$Nb & 5.256$\,\times\,10^{4}$ & 6.11\,(EC/$\beta^{+}\gamma$) & 0.024 & 0.095 \\
$^{96}$Tc & 3.698$\,\times\,10^{5}$ & 2.97\,(EC/$\beta^{+}\gamma$)& 0.012 & 0.059 \\
$^{98}$Rh & 5.232$\,\times\,10^{2}$ & 5.06\,(EC/$\beta^{+}\gamma$) & 0.011 & 0.076 \\
$^{100}$Rh & 7.488$\,\times\,10^{4}$ & 3.63\,(EC/$\beta^{+}\gamma$) & 0.088 & 0.234 \\
$^{104}$Ag & 4.152$\,\times\,10^{3}$ & 4.28\,(EC/$\beta^{+}\gamma$) & 0.012 & 0.160 \\
$^{104m1}$Ag & 2.010$\,\times\,10^{3}$ & 4.28\,(EC/$\beta^{+}\gamma$) & 0.018 & 0.111 \\
$^{107}$In & 1.944$\,\times\,10^{3}$ & 3.43\,(EC/$\beta^{+}\gamma$) & 0.019 & 0.135 \\
$^{108}$In & 3.480$\,\times\,10^{3}$ & 5.16\,(EC/$\beta^{+}\gamma$) & 0.089 & 0.194 \\
$^{110}$In & 1.771$\,\times\,10^{4}$ & 3.89\,(EC/$\beta^{+}\gamma$) & 0.053 & 0.236 \\
$^{110m1}$In & 4.146$\,\times\,10^{3}$ & 3.89\,(EC/$\beta^{+}\gamma$) & 0.066 & 0.351 \\
$^{109}$Sn & 1.080$\,\times\,10^{3}$ & 3.85\,(EC/$\beta^{+}\gamma$) & 0.027 & 0.122 \\
$^{113}$Sb & 4.002$\,\times\,10^{2}$ & 3.92\,(EC/$\beta^{+}\gamma$) & 0.036 & 0.231\\
$^{114}$Sb & 2.094$\,\times\,10^{2}$ & 5.88\,(EC/$\beta^{+}\gamma$) & 0.020 &0.297 \\
$^{115}$Sb & 1.926$\,\times\,10^{3}$ & 3.03\,(EC/$\beta^{+}\gamma$) & 0.031 & 0.839 \\
$^{116}$Sb & 9.480$\,\times\,10^{2}$ & 4.71\,(EC/$\beta^{+}\gamma$) & 0.071 & 0.939\\
$^{118}$Sb & 2.160$\,\times\,10^{2}$ & 3.66\,(EC/$\beta^{+}\gamma$) & 0.165 & 1.288 \\
$^{124}$Sb & 5.201$\,\times\,10^{6}$ & 2.90\,(EC/$\beta^{-}\gamma$) & 0.016 & 0.054\\
$^{115}$Te & 3.480$\,\times\,10^{2}$ & 4.64\,(EC/$\beta^{+}\gamma$) & 0.012 & 0.124\\
$^{117}$Te & 3.720$\,\times\,10^{3}$ & 3.54\,(EC/$\beta^{+}\gamma$) & 0.052 & 0.594\\
$^{119}$I & 1.146$\,\times\,10^{3}$ & 3.51\,(EC/$\beta^{+}\gamma$) & 0.053 & 0.533\\
$^{120}$I & 4.896$\,\times\,10^{3}$ & 5.62\,(EC/$\beta^{+}\gamma$) & 0.091 & 0.953\\
$^{122}$I & 2.178$\,\times\,10^{2}$ & 4.23\,(EC/$\beta^{+}\gamma$) & 0.289 & 1.965 \\
$^{124}$I & 3.608$\,\times\,10^{5}$ & 3.16\,(EC/$\beta^{+}\gamma$) & 0.190 & 1.654 \\
$^{130}$I & 4.450$\,\times\,10^{4}$ & 2.95\,($\beta^{-}\gamma$) & 0.195 & 1.188\\
$^{132}$I & 8.262$\,\times\,10^{3}$ & 3.58\,($\beta^{-}\gamma$) & 0.148 & 0.427\\
$^{134}$I & 3.150$\,\times\,10^{3}$ & 4.18\,($\beta^{-}\gamma$) & 0.043 & 0.183 \\
$^{121}$Xe & 2.406$\,\times\,10^{3}$ & 3.75\,(EC/$\beta^{+}\gamma$) & 0.100 & 0.540 \\
$^{125}$Cs & 2.802$\,\times\,10^{3}$ & 3.09\,(EC/$\beta^{+}\gamma$) & 0.012 & 0.266 \\
$^{126}$Cs & 9.840$\,\times\,10^{1}$ & 4.82\,(EC/$\beta^{+}\gamma$) & 0.011 & 0.080 \\
$^{128}$Cs & 2.196$\,\times\,10^{2}$ & 3.93\,(EC/$\beta^{+}\gamma$) & 0.031 & 0.229 \\
\end{tabular}
\end{ruledtabular}
\end{table}
Most of the listed isotopes are directly produced by spallation reactions.
A lager mass difference between $^{136}$Xe and the daughter particle is an indication of a greater amount of neutron emission in the xenon spallation reaction.
That is the reason we introduced neutron multiplicity into our xenon spallation identification.

\begin{figure}[h]
    \includegraphics[width=1.0\linewidth]{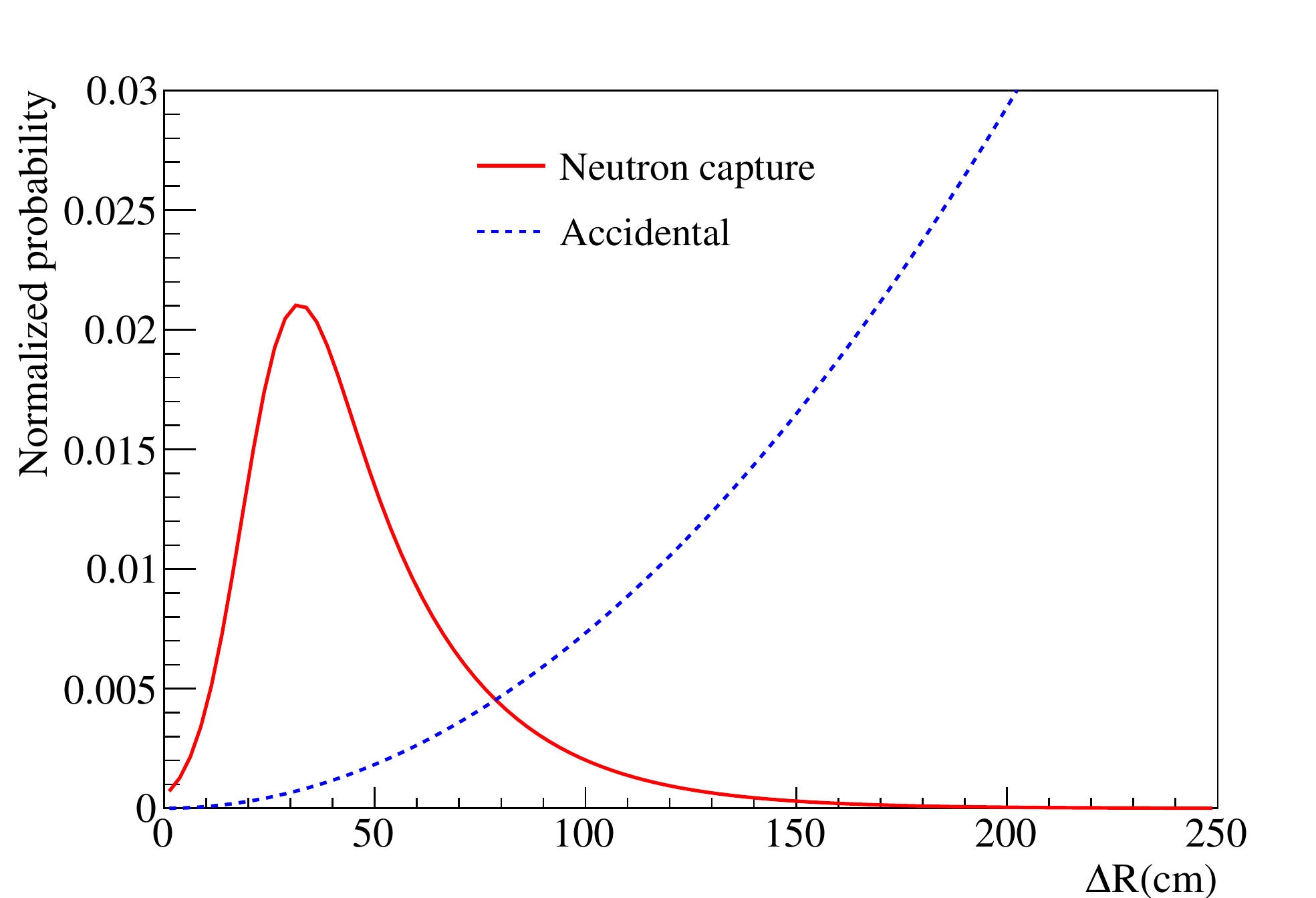}
    \caption{\label{fig:dR_PDF} PDFs of neutron capture events and accidental coincidences in Eq.~(\ref{eq:ENN}) are given as red solid line and blue dashed line, respectively.}
    \includegraphics[width=1.0\linewidth]{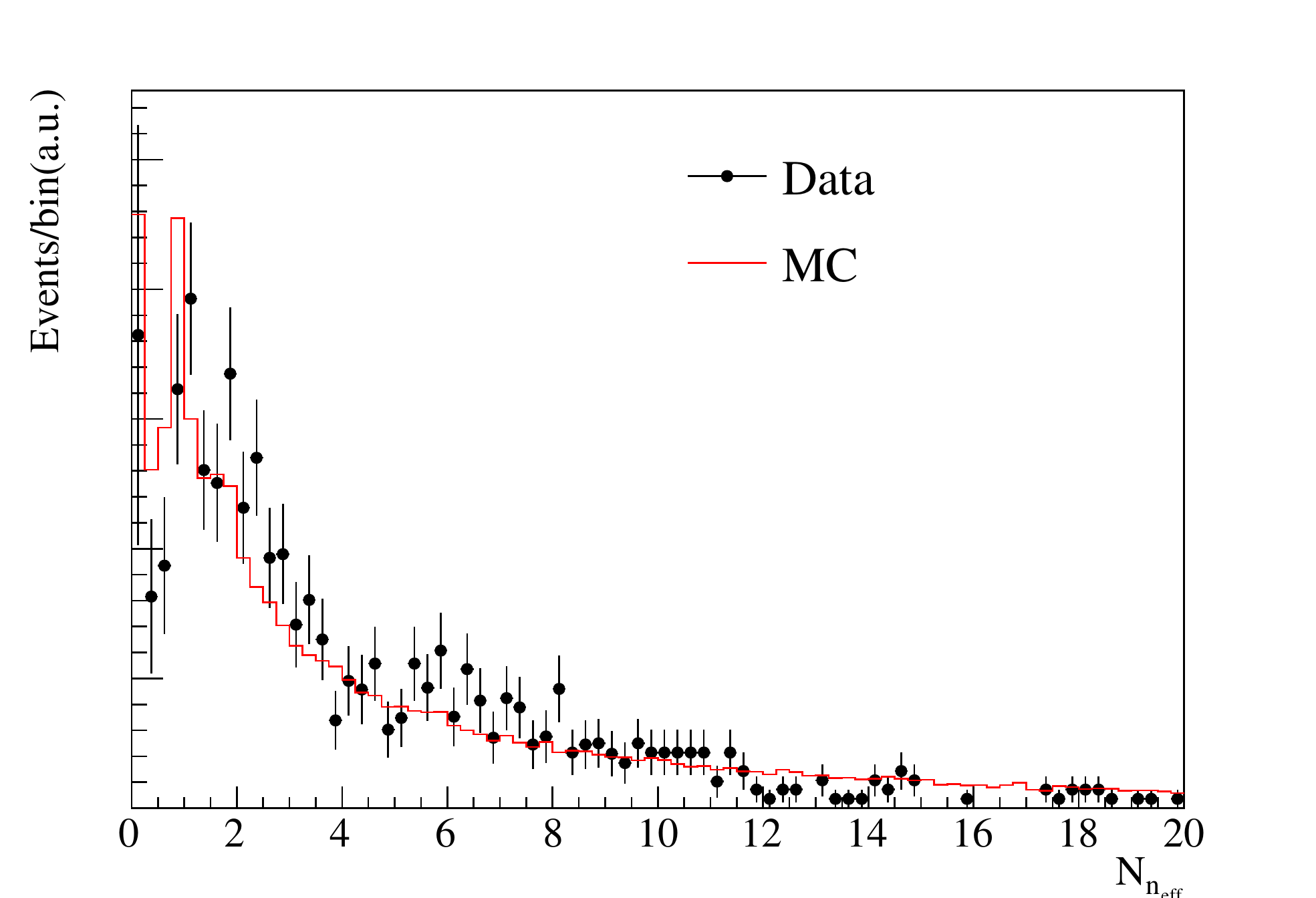}
    \caption{\label{fig:C10ENN} 
    Neutron multiplicity of $^{10}$C from simulation (red line) and data (black points) in $r<$\,4\,m with the tube cut and the hot spot veto.}
\end{figure}

The xenon spallation events are selected by a cut to the likelihood ratio,
\begin{equation}\label{eq:LL_LikelihoodRatio}
\mathcal{LR} = \frac{\mathcal{L}_{xe}}{\mathcal{L}_{xe}+\mathcal{L}_{acc}},
\end{equation}
where $\mathcal{L}_{xe}$($\mathcal{L}_{acc}$) is a likelihood of the xenon spallation (accidentals) constructed as a function of $\Delta T$, $\Delta R_{min}$ and the effective number of neutron captures ($N_{n_{eff}}$).
$\Delta T$ and $\Delta R_{min}$ are introduced in Sec.~\ref{sec:NeutronCapture} and \ref{sec:carbon}, while $N_{n_{eff}}$ is defined as follows.
\begin{equation}\label{eq:ENN}
    N_{n_{eff}} \equiv \sum_{j} \frac{P_{n}(\Delta R_{j})}{P_{n}(\Delta R_{j}) + P_{acc}(\Delta R_{j})}, 
\end{equation}
where $\Delta R_{j}$ is the distance between a $\beta$ decay and the $j$-th neutron capture, $P_{n}$ and $P_{acc}$ are PDFs of neutron capture events and accidental coincidences (see Fig.~\ref{fig:dR_PDF}, \ref{fig:C10ENN}).
$\mathcal{L}_{xe}$ is calculated by iterating over the $i$-th daughter nucleus:
\begin{eqnarray}\label{eq:LL_Likelihood}
    \mathcal{L}_{xe} &=& \sum_{i} P_{xe_{i}}(\Delta R_{min}, N_{n_{eff}}, \Delta T),\\
    \mathcal{L}_{acc} &=& P_{acc}(\Delta R_{min}, N_{n_{eff}}, \Delta T),
\end{eqnarray}
where the likelihood PDFs of xenon spallation ($P_{xe_i}$) are made assuming the production ratio of daughters follow the FLUKA prediction, while accidental PDFs ($P_{acc}$) to be uniform in time and space.
Selected events with $\Delta T<4\times10^{5}$\,s are vetoed, generating $\sim$\,9\% of dead-time.
The choice of the veto time is based on the trade off between veto efficiency and dead-time.
The selection efficiency is evaluated to be 42.0\,$\pm$\,8.8\%.

\begin{figure}[h]
    \includegraphics[width=1.0\linewidth]{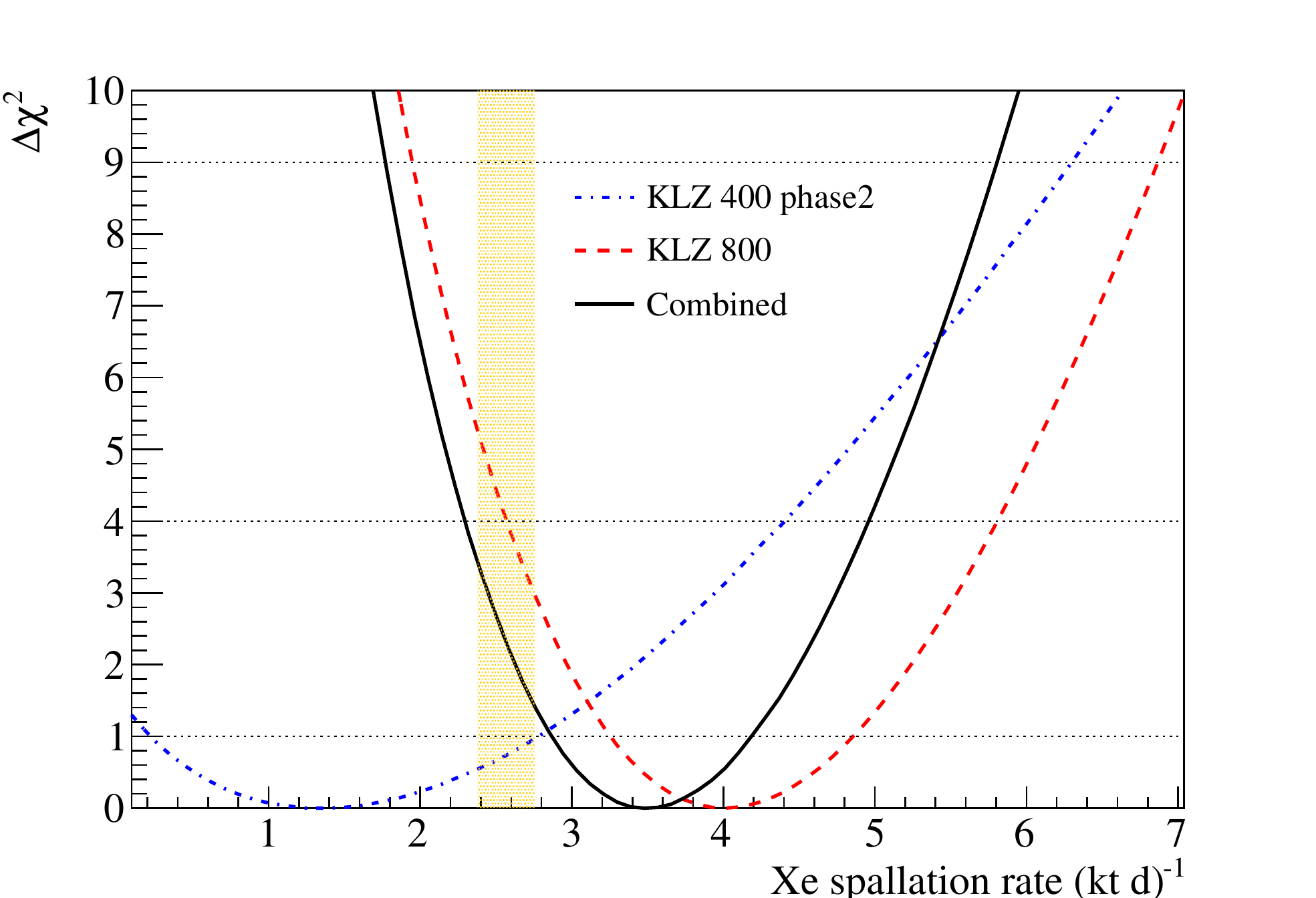}
    \caption{\label{fig:LLrate} $\Delta\chi^2$ of the xenon spallation rate fitted to the energy spectrum. 
    The results of KamLAND-Zen 400 phase2 and KamLAND-Zen 800 are shown as blue and red dashed lines, and the combined result is shown in black. The FLUKA prediction is indicated with a orange band.}
\end{figure}

In the $0\nu\beta\beta$ decay search, we estimate the $0\nu\beta\beta$ decay rate and xenon spallation rate by the fit to the energy spectrum.
The background models and vetoes are detailed in \cite{Zen800}.
After applying the vetoes and the xenon spallation likelihood selection to the $0\nu\beta\beta$ decay candidates, the xenon spallation production rate is evaluated from a Poisson-$\chi^2$ scan as shown in  Fig.~\ref{fig:LLrate}.
We observe a xenon spallation rate of 3.5\,$\pm$\,0.6\,(kton\,day)$^{-1}$ in 2.35$\,\leq\,E\,\leq\,$2.70\,MeV, while the simulation prediction is 2.6\,$\pm$\,0.2\,(kton\,day)$^{-1}$.

\begin{figure}[h]
    \includegraphics[width=1.0\linewidth]{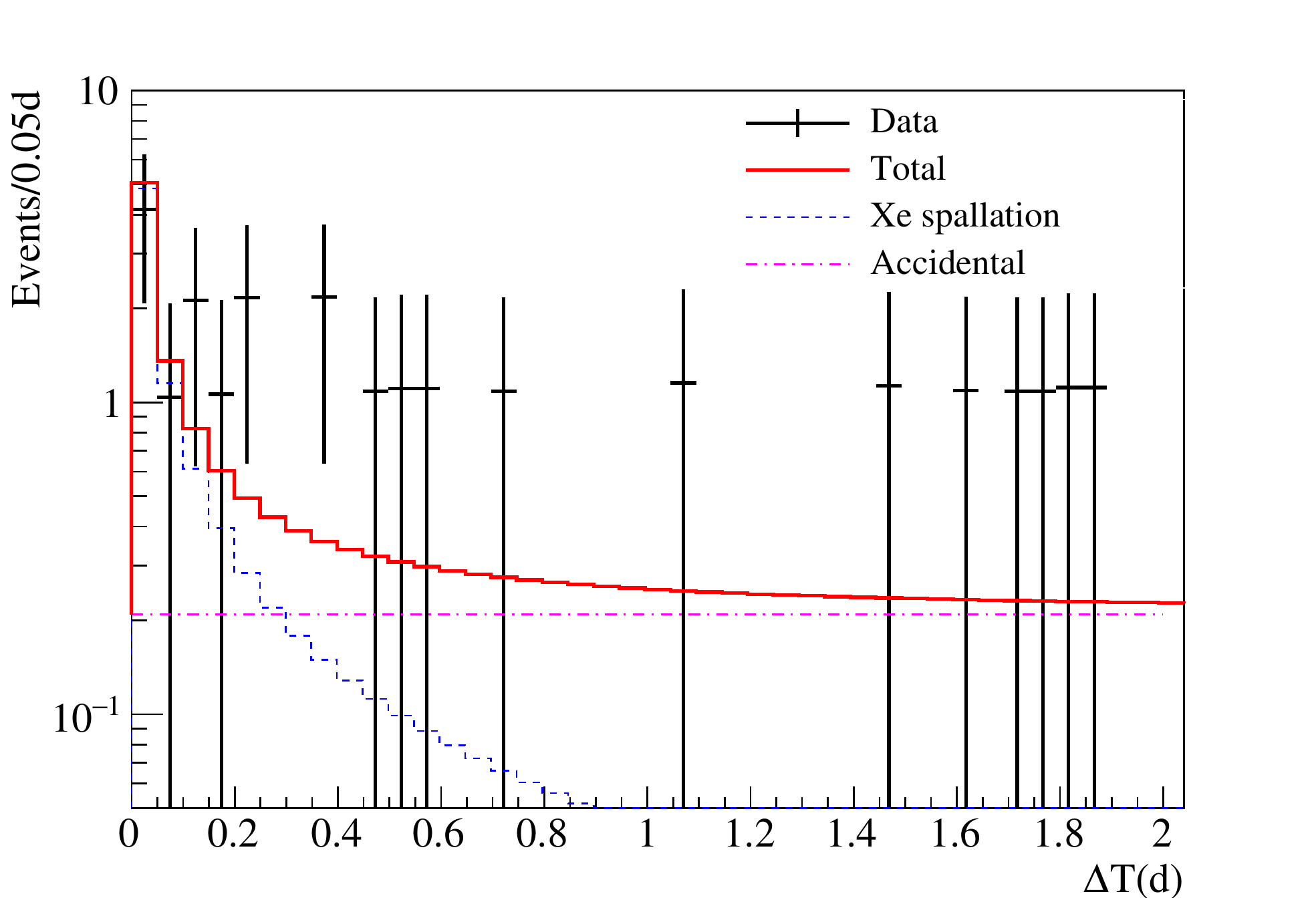}
    \caption{\label{fig:LLdT}Time difference between the cosmic-ray muon and xenon spallation candidates. The data was plotted after correcting the muon detection efficiency.}
\end{figure}

The time difference between the cosmic-ray muon and xenon spallation candidates provides another estimate for the xenon spallation production (Fig.~\ref{fig:LLdT}).
Xenon spallation candidates in $2.4<E<3.0$\,MeV and $r<1.6$\,m with the hot spot veto are selected, where the detector livetime is 625\,days and Xe-LS mass 13\,ton.
In order to discriminate the spallation events other than xenon spallation, (a) a 150-ms-veto, (b) a 180-s-veto and (c) a 27-min-veto were applied after the muon, where (b) and (c) require three-fold tagged events.
Short lifetime spallation events are removed by (a).
$^{10}$C $\beta^{+}$ decay is discriminated by (b) 
and $^{137}$Xe $\beta^{-}$ decay is suppressed using (c) in combination with the requirement of more than one emitted neutron with $N_{S}>240$.

Accidental coincidences are discriminated by requiring $N_{n_{eff}}\ge6$.
We assume that accidental coincidences follow a Poisson distribution with an average event rate of 3.6\,/day.
The number of observed events in $0<\Delta T<1$\,day is 16, which deviates from the accidental only model by 4.8\,$\sigma$.
Poisson-$\chi^2$ is calculated as a function of the production ratio of measurement to simulation. 
The ratio was estimated at 1.27\,$^{+\,0.47}_{-0.4}$.
 
\subsection{\label{sec:ncaptureXe136}Neutron capture on $^{136}$Xe}
The neutron capture reaction on $^{136}$Xe is followed by $^{137}$Xe $\beta^{-}$ decay ($\tau_{1/2}$\,=\,3.8\,min, $Q\,=4.173\,$MeV):
\begin{itemize}
    \item[(1)] $^{136}$Xe + $n \,\rightarrow\, ^{137}$Xe + $\gamma$
    \item[(2)] $^{137}$Xe $\rightarrow\, ^{137}$Cs + $\beta^{-} + \bar{\nu_e} + \gamma$
\end{itemize}

Using the neutron capture ratio of $^{136}$Xe, according to Table~\ref{tab:CaptureRatio}, and the measured neutron capture rate in the Xe-LS, see Table~\ref{tab:nCapOn1H}, the event rate of neutron capture on $^{136}$Xe is expected to be: 
\begin{eqnarray}\label{eq:Xe137exp}
\mathcal{R}_{\rm{^{137}Xe}} &=& 4347 \times 0.00107 \notag\\
&=& 4.6\,\pm\,0.5\,\rm{(kton\,day)^{-1}}, \notag  
\end{eqnarray}
while the FLUKA prediction is  5.5\,$\pm$\,0.4\,(kton\,day)$^{-1}$.

\begin{figure}[h]
    \includegraphics[width=1.0\linewidth]{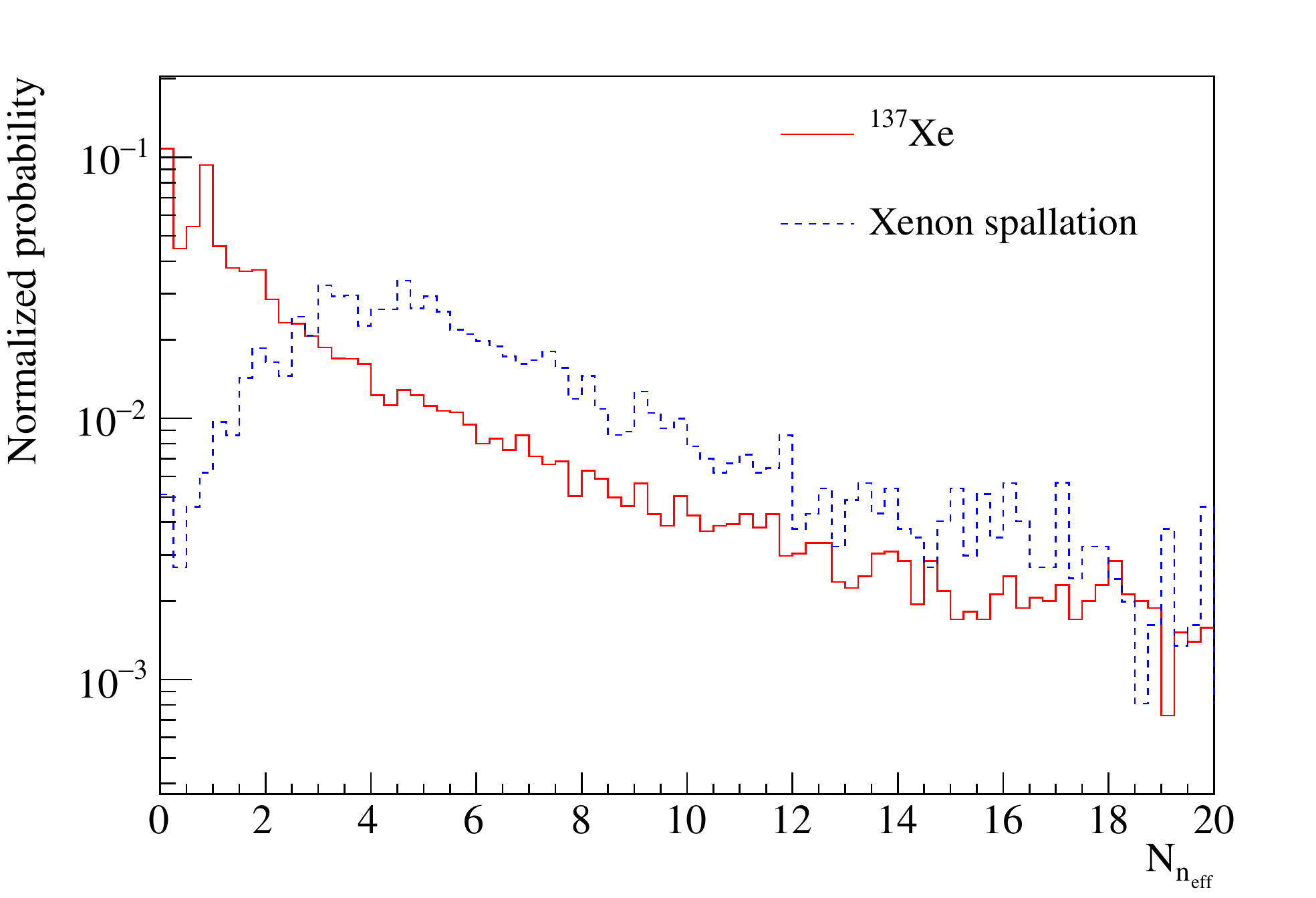}
    \includegraphics[width=1.0\linewidth]{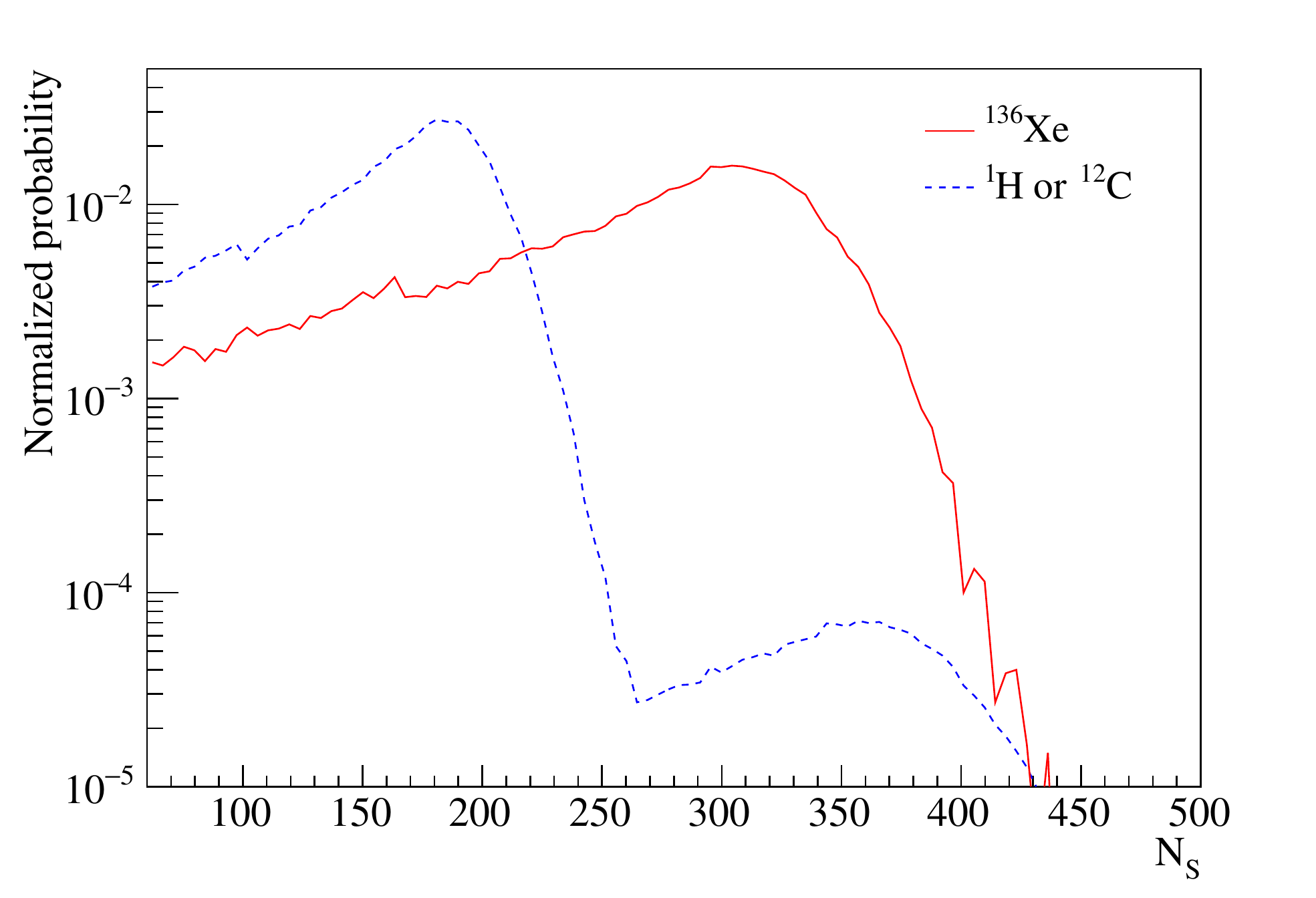}
    \caption{\label{fig:PDFs} Upper figure shows the PDFs of neutron multiplicity of $^{137}$Xe events (red solid) and other xenon spallation events (blue dashed). 
    Bottom figure shows the PDFs of $N_{S}$ for neutron captures on $^{136}$Xe (red solid) and $^{1}$H and $^{12}$C (blue dashed).}
\end{figure}

We identify $^{137}$Xe $\beta^{-}$ decay by a triple coincidence of a LS muon, the $\gamma$-ray from neutron capture ($E_\gamma = 4.025\,$MeV) and the subsequent $\beta^{-}$ decay.
The $\gamma$-ray and the $\beta^{-}$ decay are selected by a likelihood,
\begin{eqnarray}\label{eq:xe_Likelihood}
    \mathcal{L} = \max(P(\Delta R)\cdot P(N_S)) \cdot P(N_{n_{eff}}),
\end{eqnarray}
where $\Delta R$, $N_S$ and $N_{n_{eff}}$ are defined in previous sections and the PDFs are given in Fig.~\ref{fig:dR_PDF} and Fig.~\ref{fig:PDFs}.
$P(\Delta R) \cdot P(N_{S})$ is calculated for each neutron capture and search for a neutron close to the $\beta$ decay with high $N_{S}$.
The dominant backgrounds for the measurement of neutron capture on $^{136}$Xe are carbon and xenon spallation, two-neutrino double-beta decay and radioactive impurities.
The $^{11}$C $\beta^{+}$ decay and two-neutrino double-beta decay rates are negligible in 2.4$\,\leq\,E\,<\,3.2$\,MeV, while 
$^{12}$B $\beta^{-}$ decay is rejected by a 150\,ms-veto after a cosmic-ray muon. 
Other carbon spallation backgrounds, with $^{10}$C $\beta^+$ decay the most critical, are suppressed by vetoing the events which are selected by n-tag in Sec.~\ref{sec:carbon} and observed within 80\,s from the cosmic-ray muon. 
The xenon spallation backgrounds and radioactive impurities are suppressed by a cut to the likelihood ratio. 

\begin{figure}[h]
    \includegraphics[width=1.0\linewidth]{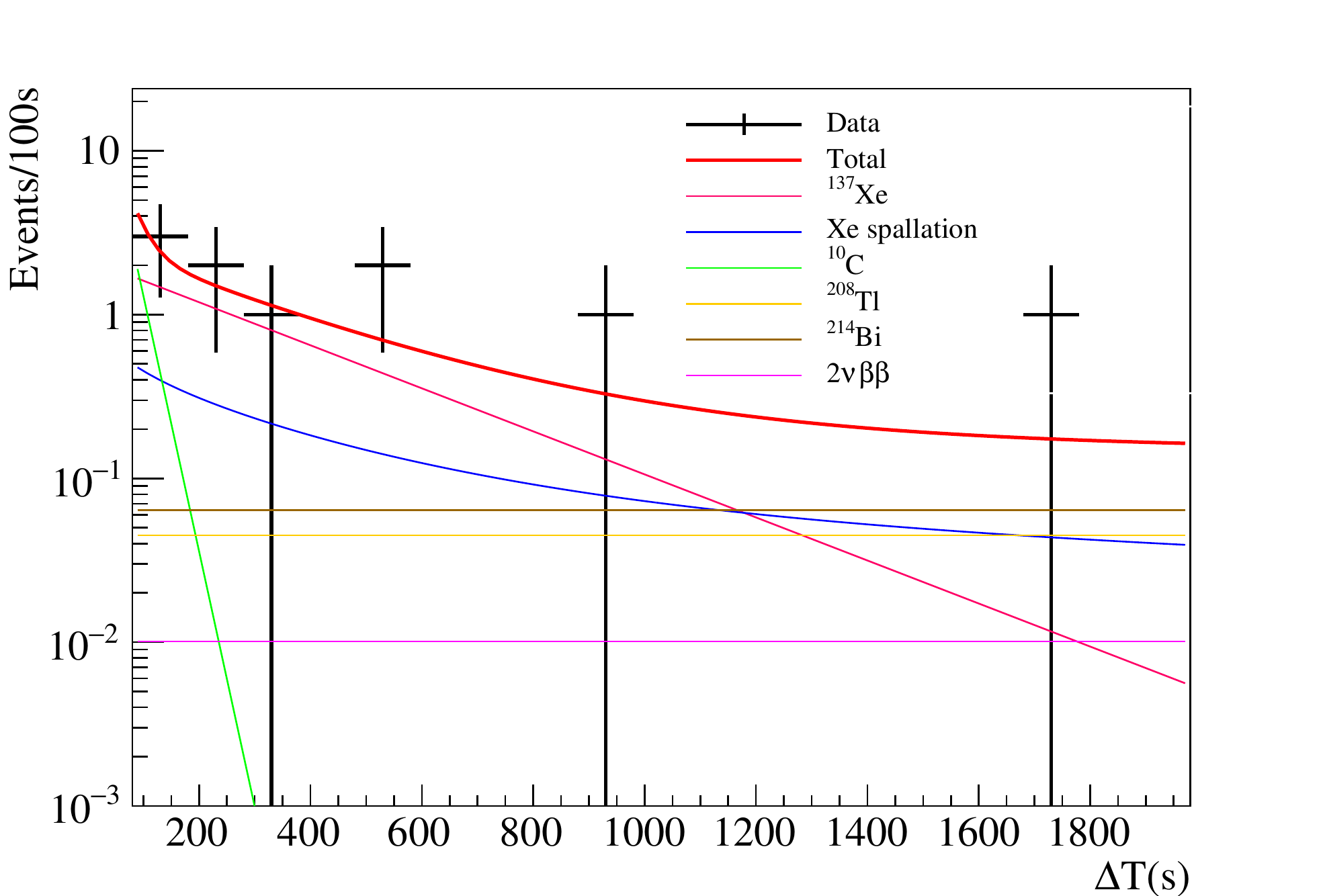}
    \caption{\label{fig:Xe137dTfit} Time difference between a cosmic-ray muon and the $^{137}$Xe $\beta^{-}$ decay candidate in $2.4\leq\,E\,<3.2$\,MeV. $^{208}$Tl $\beta^{-}$ decay is derived from the $^{232}$Th series.}
    \includegraphics[width=1.0\linewidth]{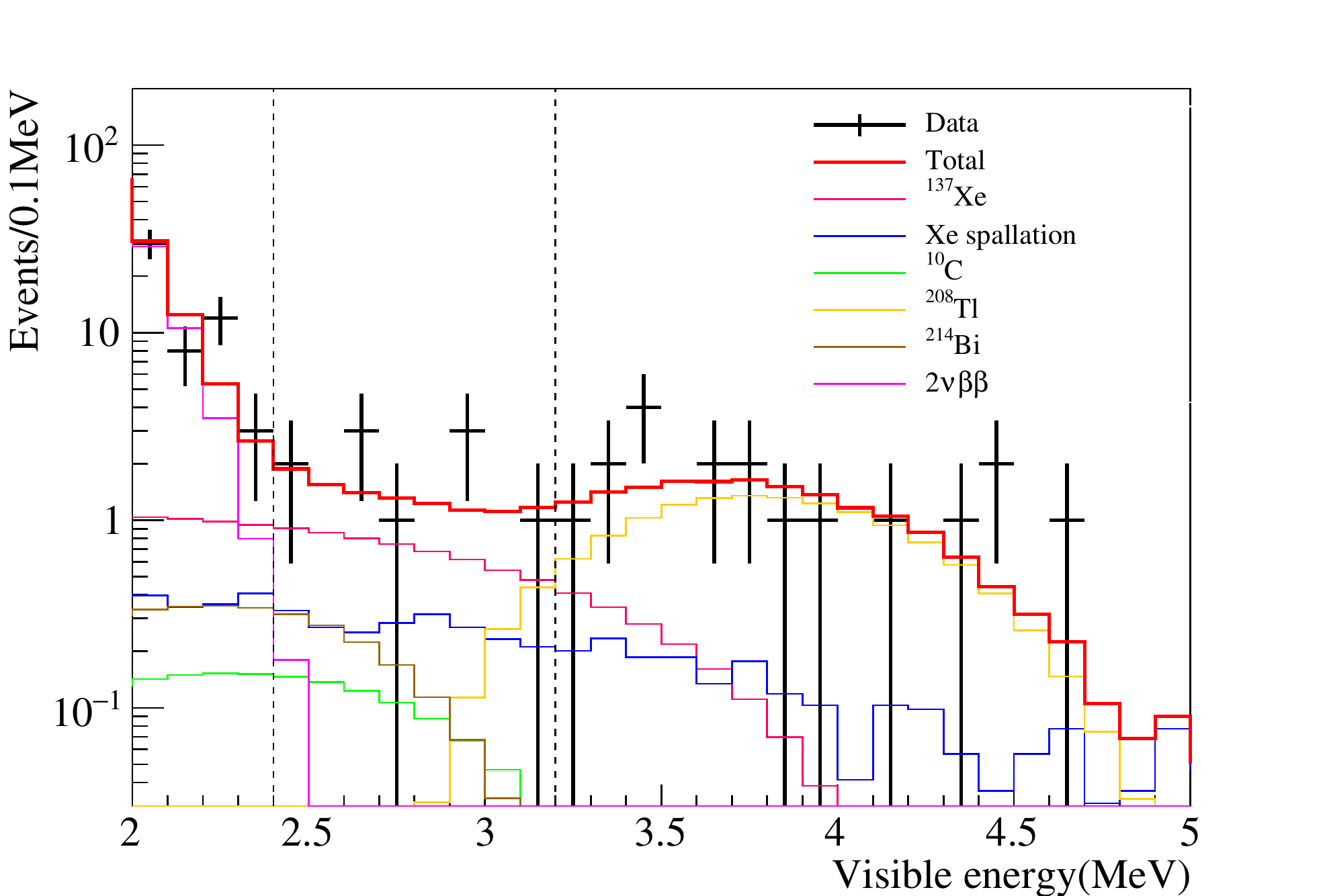}
    \caption{\label{fig:Xe137Efit} Visible energy of the $^{137}$Xe $\beta^{-}$ decay candidates in $80\,\leq\,\Delta T\,<\,1980$\,s. The fit range is indicated by black dashed lines.}
\end{figure}

Similar to the measurement of carbon spallation production in Eq.(\ref{eq:Cfit}), the capture rate is obtained from a simultaneous fit to time and energy.
Figure~\ref{fig:Xe137dTfit} shows the time difference between a cosmic-ray muon and the $^{137}$Xe $\beta^{-}$ decay candidate which is located within 1.9\,m from the detector center. 
The detector livetime is 646\,days and the Xe-LS mass is 22\,ton. 
The visible energy spectrum of the $^{137}$Xe $\beta^{-}$ decay candidates in  $80\,\leq\,\Delta T\,<\,1980$\,s is shown in Fig.~\ref{fig:Xe137Efit}.
The event rate of $^{10}$C $\beta^{+}$ decay and xenon spallation are constraint from the measurements in Sec.~\ref{sec:carbon} and Sec.~\ref{sec:xenon}.
The number of events is evaluated from a Poisson-$\chi^{2}$ scan.
We observe 4.6$\,\pm\,$2.8\,(stat.)$\,\pm\,0.2$\,(syst.)\,(kton\,day)$^{-1}$ of neutron captures on $^{136}$Xe, corresponding to 236\,$\pm$\,145\,mb.
The measurement is consistent with the expectation in Eq.~(\ref{eq:Xe137exp}) using the $^{136}$Xe capture cross section of ~\cite{Xecapture}.

\section{\label{sec:summary}Summary}
The radioisotopes production caused by cosmic-ray muon spallation in a xenon-loaded liquid scintillator was measured with KamLAND and compared to results from the FLUKA and Geant4 simulation codes.
The production yield of carbon spallation isotopes is consistent with our previous measurement~\cite{spallation2010} and additionally the $^{6}$He production rate is measured for the first time.
These measurements were done using a combination of delayed coincidence with a higher neutron detection efficiency and the shower likelihood method to cover the contribution from non-neutron emitting reactions.

The xenon spallation productions including their subsequent decays were studied with FLUKA and Geant4 simulations.
Figure~\ref{fig:EnergySpectra} and Table~\ref{tab:criticalLonglived} show the non-negligible isotopes for the $0\nu\beta\beta$ decay search.
Their contributions in the ROI is estimated to be $2.6\,\pm\,$0.2\,(kton\,day)$^{-1}$.

We developed a likelihood method which effectively utilizes the space correlation of muon-induced neutron captures and their multiplicity, resulting in a xenon spallation selection efficiency of $42.0\,\pm\,8.8$\%.
The observed amount of xenon spallation production is $3.5\,\pm\,0.6\,$(kton\,day)$^{-1}$ in $2.35\,\leq\,E\,\leq\,2.70\,$MeV, an important background for the $0\nu\beta\beta$ decay search.
One of the possible approaches to refine the background discrimination is through particle identification, since most of the xenon spallation decays are accompanied by $\gamma$-rays.

The $^{137}$Xe $\beta^{-}$ decay is an inevitable background for the $0\nu\beta\beta$ search with $^{136}$Xe.
We measured the production yield using a two-dimensional binned likelihood fit.
Although there are large statistical uncertainties, a significant amount of neutron capture on $^{136}$Xe is observed.
The observation is consistent with the expectation from \cite{Xecapture}.

\begin{acknowledgments}
The KamLAND-Zen experiment is supported by JSPS KAKENHI Grant Numbers 21000001, 26104002, and 19H05803; the Dutch Research Council (NWO); and under the U.S. Department of Energy (DOE) Grant No.De-AC02-05CH11231, other DOE, and NSF grants to individual institutions. 
We appreciate the supports of the Kamioka Mining and Smelting Company for activities in the mine, and NII for SINET4 as well.
\end{acknowledgments}


\end{document}